\documentclass[lettersize,journal]{IEEEtran}
\usepackage{amsmath,amsfonts}
\usepackage{algorithmic}
\usepackage{algorithm}
\usepackage{array}
\usepackage[caption=false,font=normalsize,labelfont=sf,textfont=sf]{subfig}
\usepackage{textcomp}
\usepackage{stfloats}
\usepackage{url}
\usepackage{verbatim}
\usepackage{graphicx}
\usepackage{cite}

\newcommand{\rfig}[1]{Fig.\,\ref{#1}}

\usepackage{bm}
\usepackage{color}
\usepackage{amsfonts}
\usepackage{mathtools}
\usepackage{upgreek}
\usepackage{fancybox}
\usepackage{here}
\usepackage{mathrsfs}

\usepackage{comment}
\usepackage{amsmath,amssymb,amsfonts}
\usepackage{amssymb}
\usepackage{dsfont}
\usepackage{textcomp}
\usepackage{xcolor}
\usepackage{lscape}
\usepackage{multicol}
\usepackage[hang,small,bf]{caption}
\usepackage[subrefformat=parens]{subcaption}
\usepackage{xcolor}
\usepackage{multirow}
\usepackage{makecell}
\usepackage{stfloats}

\begin{document}

\title{Clustering-based Recurrent Neural Network Controller synthesis under Signal Temporal Logic Specifications}

\author{
Kazunobu Serizawa,~\IEEEmembership{Graduate Student Member,~IEEE,}
~Kazumune Hashimoto,~\IEEEmembership{Member,~IEEE,}

~Wataru Hashimoto,
~Masako Kishida,~\IEEEmembership{Senior Member,~IEEE,}
~Shigemasa Takai,~\IEEEmembership{Senior Member,~IEEE}
        % <-this % stops a space
\thanks{Kazunobu Serizawa, Kazumune Hashimoto, Wataru Hashimoto, and Shigemasa Takai are with graduate school of Engineering, The University of Osaka, Osaka, Japan (E-mail address: serizawa.kazunobu.ogp@ecs.osaka-u.ac.jp, hashimoto@eei.eng.osaka-u.ac.jp, hashimoto@is.eei.eng.osaka-u.ac.jp, takai@eei.eng.osaka-u.ac.jp).
Masako Kishida is with the National Institute of Informatics, Tokyo, Japan (E-mail address: kishida@nii.ac.jp).
This work is partially supported by JST CREST JPMJCR201, JST ACT-X JPMJAX23CK, and JSPS KAKENHI Grant 21K14184, and 22KK0155.}% <-this % stops a space
\thanks{Manuscript submitted April 18, 2025.}}

%\markboth{IEEE TRANSACTIONS ON CONTROL SYSTEMS TECHNOLOGY}%


\maketitle

\begin{abstract}
    Autonomous robotic systems require advanced control frameworks to achieve complex temporal objectives that extend beyond conventional stability and trajectory tracking. Signal Temporal Logic (STL) provides a formal framework for specifying such objectives, with robustness metrics widely employed for control synthesis. 
    Existing optimization-based approaches using neural network (NN)-based controllers often rely on a single NN for both learning and control. 
    However, variations in initial states and obstacle configurations can lead to discontinuous changes in the optimization solution, thereby degrading generalization and control performance. To address this issue, this study proposes a method to enhance recurrent neural network (RNN)-based control by clustering solution trajectories that satisfy STL specifications under diverse initial conditions. The proposed approach utilizes trajectory similarity metrics to generate clustering labels, which are subsequently used to train a classification network. This network assigns new initial states and obstacle configurations to the appropriate cluster, enabling the selection of a specialized controller. By explicitly accounting for variations in solution trajectories, the proposed method improves both estimation accuracy and control performance. Numerical experiments on a dynamic vehicle path planning problem demonstrate the effectiveness of the approach.
\end{abstract}


\section{Introduction}
\IEEEPARstart{C}{yber-Physical} Systems (CPSs) integrate computational (cyber) and physical components, enabling real-time interactions between digital and physical domains. These systems typically comprise sensors, actuators, controllers, and communication networks that monitor and regulate physical processes.  A notable subclass of CPSs, autonomous robotic systems—such as drones, robots, and self-driving cars—are designed to perform critical mission tasks that involve complex temporal objectives, including periodic, sequential, and reactive behaviors. These objectives extend beyond conventional control goals, such as stability and trajectory tracking, necessitating more sophisticated specification frameworks.

To address these challenges, researchers have developed formal specification languages that provide a rigorous framework for defining temporal and spatial constraints in autonomous systems. A prominent example is Signal Temporal Logic (STL) \cite{maler2004monitoring}, which is particularly suited for reasoning about continuous real-valued signals. STL introduces a quantitative measure known as robustness \cite{donze2010robust}, which quantifies the degree to which a given formula is satisfied. This robustness metric has been widely applied in control systems \cite{raman2014,Sadraddini2015,pant2017smooth,takayama2023,leung2023backpropagation,lindemann2019control,yu2024continuous,verginis2024planning}. 
For instance, \cite{raman2014} formulates STL constraints and robustness measures as mixed-integer linear constraints, which are then integrated into Model Predictive Control (MPC). To mitigate the computational complexity associated with mixed-integer programming, subsequent studies \cite{pant2017smooth} introduced smooth robustness, enabling efficient robustness maximization through gradient-based optimization techniques. 
In \cite{takayama2023}, the authors propose leveraging a structure-aware decomposition of STL formulas to reformulate the original problem as a difference of convex (DC) programs. This reformulation facilitates the application of the convex-concave procedure (CCP), allowing the problem to be solved sequentially through a series of convex quadratic programs. 
 
More recently, neural networks (NNs) have emerged as a powerful tool for synthesizing controllers that satisfy STL specifications. In \cite{liu2021recurrent}, recurrent neural networks (RNNs) are employed to develop control policies that capture the temporal dependencies inherent in STL constraints. Another approach \cite{yaghoubi2019worst} integrates feedforward NN controllers with the Lagrange multiplier method to maximize STL robustness. Additionally, \cite{stlmpc} introduces a NN-based controller synthesized under STL constraints, akin to MPC, where the trained controller predicts a trajectory within a planning horizon to ensure STL satisfaction. To enhance safety, a backup policy is incorporated to handle scenarios in which the learned controller fails to generate STL-compliant trajectories. 
In \cite{PuranicSTL}, the authors propose leveraging the STL formulae to evaluate and rank the quality of demonstrations, facilitating the inference of reward functions that guide reinforcement learning algorithms. In another line of research, \cite{leung2023backpropagation} proposed to translate STL robustness formulas into computation graphs, referred to as STLCG, enabling the use of automatic differentiation tools to backpropagate through these formulas. This integration allows for the seamless incorporation of STL specifications into various gradient-based approaches, enhancing the generalization capabilities of learned control policies.
Furthermore, \cite{hashimoto2022stl2vec} presents STL2vec, a method that encodes STL specifications as vector representations, enabling controllers to accommodate multiple specifications effectively.
\IEEEpubidadjcol

A common limitation of the aforementioned approaches is their reliance on a \textit{single} NN controller for both learning and control. 
However, even when the control objective (i.e., STL specification) remains unchanged, the solution to the optimization problem may exhibit significant discontinuities due to variations in environmental factors, such as the system’s initial state and the position or shape of obstacles. Consequently, enhancing the generalization capability of a single controller to handle a wide range of initial states and design parameters substantially increases the complexity of the learning process and may degrade both accuracy and control performance.

To address this issue, this study proposes a method to improve the control performance of an RNN-based controller by \textit{clustering} solution trajectories that satisfy a given STL specification under randomly generated initial states and obstacle parameters.
%再度確認する
%\color{blue}
%As shown in \cite{patil2022robustness, Narendra1997adaptive, chen2001nonliner}, switching between multiple controllers can improve control performance. Motivated by these findings, the proposed method introduces a classification network to assign appropriate controllers based on initial conditions and obstacle parameters.
%\color{black}
First, for various initial states and obstacle configurations, optimal solution trajectories obtained by solving an optimal control problem are transformed into feature vectors, which are subsequently used to generate clustering labels. In this transformation process, cumulative errors between state sequences are utilized to quantitatively assess trajectory similarity. Next, a neural network, referred to as the classification network, is trained to classify any given initial state and obstacle configuration into the appropriate cluster. Finally, based on the classification network’s output—i.e., the predicted cluster corresponding to the initial state and obstacle parameters—the RNN-based controller associated with the identified cluster is assigned. By explicitly accounting for variations in solution trajectories with respect to initial state and obstacle changes, the proposed method aims to improve both the accuracy and control performance of the learned controller. 

In summary, the main contribution of this paper is summarized as follows. 
\begin{itemize}
    \item We introduce a novel approach for synthesizing RNN-based controllers under STL specifications. Unlike conventional methods that rely on a single NN controller, the proposed framework clusters solution trajectories based on variations in initial states and obstacle configurations. This clustering enables the selection of specialized controllers, thereby improving control performance compared with a single NN controller.
    \item We propose an additional NN, referred to as a {classification network}, trained to assign new initial states and obstacle parameters to the appropriate cluster. This allows for the adaptive selection of specialized controllers. By leveraging trajectory clustering and classification, our approach enhances robustness and improves the generalization capability of the learned control policies. 
\end{itemize}
Finally, the effectiveness of the proposed approach is validated through numerical experiments on a dynamic vehicle path planning problem.

\section{Preliminaries of Signal Temporal Logic}\label{STL}
In this section, we summarize the fundamentals of the STL \cite{maler2004monitoring}. 
The \textit{syntax} of STL is constructed using basic propositions, logical connectives, and temporal operators. STL formulas are defined recursively as follows: 
\begin{align}\label{grammar}
    \phi ::= &\top \mid \mu \mid \neg \phi \mid \phi_1 \land \phi_2 \mid \phi_1 \lor \phi_2 \mid  \phi_1 \bm{U}_{I} \phi_2 
\end{align} 
where $\mu : \mathbb{R}^n \rightarrow \mathbb{B}$ is the predicate whose boolean truth value is determined by the sign of a function $h : \mathbb{R}^n \rightarrow \mathbb{R}$ of an $n$-dimensional vector $x$, i.e., $\mu$ is true if $h(x)>0$, and false otherwise. $\phi$, $\phi_1$, and $\phi_2$ represent STL formulae, $\top$, $\neg$, $\land$, and $\lor$ are Boolean \textit{true}, \textit{negation}, \textit{and}, and \textit{or} operators, respectively, and $\bm{U}_{I}$ is the temporal \textit{until} operator defined on a time interval $I =[a, b] = \{ t \in \mathbb{Z}_{\geq 0}: a\leq t \leq b\}$ ($a$, $b\in \mathbb{Z}_{\geq 0}$).

The Boolean semantics of STL describe how the truth value of an STL formula is evaluated over a signal $x_{k}$, $k \in \mathbb{N}$. 
More precisely, the {Boolean} semantics of $\phi$ with respect to the trajectory over a given horizon $T$ from time $t$, i.e., $x_{t:T} := x_t, ..., x_T$, is defined as follows:  
\begin{align}
    &x_{t:T} \models \mu \Leftrightarrow h (x_t)>0\notag \\
    &x_{t:T}\models \neg \mu \Leftrightarrow \neg (x_{t:T}\models \mu) \notag\\
    &x_{t:T}\models \phi_1 \land \phi_2 \Leftrightarrow x_{t:T}\models \phi_1 \land x_{t:T}\models \phi_2 \notag \\
    &x_{t:T}\models \phi_1 \lor \phi_2 \Leftrightarrow x_{t:T} \models \phi_1 \lor x_{t:T} \models \phi_2 \notag \\
     & x_{t:T}\models \phi_1 \bm{U}_{I}\phi_2 \Leftrightarrow \exists t_1 \in t+I\  \mathrm{s.t.}\  x_{t_1:T}\models \phi_2 \notag \\
     & \qquad \qquad \land \forall t_2 \in [t,t_1], x_{t_2:T}\models \phi_1\notag,
\end{align}
where $t+ I = \{t+k \in \mathbb{Z}_{\geq 0}: k \in I \}$.  
Intuitively, $\phi_1 \bm{U}_{I} \phi_2$ means that ``$\phi_2$ holds for the signal within a time interval $I$ and $\phi_1$ must always be true against the signal prior to that". 
Other temporal operators $eventually$ and $always$ ($\bm{F}_{I}$ and $\bm{G}_{I}$) are defined based on the until operator as $\bm{F}_{I}\phi:= \top \bm{U}_{I}\phi$ and $\bm{G}_{I}\phi:=\neg \bm{F}_{I}\neg \phi$ respectively.  $\bm{F}_{I}\phi$ states that ``$\phi$ must hold at some time point within the interval $I$" while $\bm{G}_{I}\phi$ states that ``$\phi$ must hold for the signal within $I$". 

The concept of robustness in STL extends the Boolean semantics by providing quantitative semantics that measure the degree to which a trajectory satisfies or violates a given STL formula. Robustness is defined such that a positive value indicates satisfaction, while a negative value indicates violation. The robustness score of an STL formula $\phi$ over a trajectory $x_{t:T}$ is inductively defined as follows: 
\begin{align}
    & \rho^\mu(x_{t:T}) = h (x_t) \notag \\
    & \rho^{\neg \mu}(x_{t:T}) = -h (x_t)\notag \\ 
    & \rho^{\phi_1 \land \phi_2}(x_{t:T}) = \min (\rho^{\phi_1} (x_{t:T}), \rho^{\phi_2} (x_{t:T}))\notag \\
    & \rho^{\phi_1 \lor \phi_2}(x_{t:T}) = \max (\rho^{\phi_1} (x_{t:T}), \rho^{\phi_2} (x_{t:T}))\notag \\
     & \rho^{\phi_1 \bm{U}_{I}\phi_2}(x_{t:T}) = \max_{t_1\in t+I}\Bigl(\min (\rho^{\phi_2}(x_{t_1:T}), \notag  \\
    & \qquad \qquad \qquad \quad \min_{t_2 \in [t, t_1]}\rho^{\phi_1}(x_{t_2:T}))\Bigr)\notag \notag 
\end{align}
Note that the trajectory horizon $T$ must be sufficiently large to appropriately capture the robustness score for the given trajectory. 
 
\section{Problem setup}\label{sec:problemsetup}
\subsection{Dynamics and environment}
We consider a discrete-time dynamical system given by  
\begin{align}
x_{k+1} = f(x_k, u_k), \  x_0 \in \mathcal{X}_0, \  u_k \in \mathcal{U}, \  k \in \mathbb{N}, \label{dynamics}
\end{align}  
where \( k \in \mathbb{N} \) denotes the time step, \( x_k \in \mathcal{X} \subset \mathbb{R}^{n_x} \) represents the system state at time \( k \), and \( u_k \in \mathcal{U} \subset \mathbb{R}^{n_u} \) is the control input. The state space is denoted by \( \mathcal{X} \), while the control input space is given by \( \mathcal{U} \). The initial state \( x_0 \) is drawn from a set \( \mathcal{X}_0 \subset \mathcal{X} \), and the control input \( u_k \) is constrained within a hyper-rectangle $\mathcal{U} = [u_{\min}, u_{\max}]$, where \( u_{\min} = [u_{\min,1}, \dots, u_{\min,n_u}]^\top \) and \( u_{\max} = [u_{\max,1}, \dots, u_{\max,n_u}]^\top \) represent the lower and upper bounds, respectively.  
The initial state \( x_0 \) is assumed to be sampled from a probability distribution \( p_{x_0}(\cdot) \) with support in \( \mathcal{X}_0 \). 
Given an initial state \(x_0 \in \mathcal{X}_0\) and a sequence of control inputs  \( u_{0:T-1} = (u_0, u_1, \dots, u_{T-1}) \), the corresponding state sequence 
  \( x_{1:T} = (x_1, x_2, \dots, x_T) \) is generated according to the dynamics in \eqref{dynamics}.
Throughout this paper, we refer to \( x_{0:T} = (x_0, x_1, \dots, x_T) \), which includes the initial state and the corresponding state sequence, 
as the \textit{state trajectory}, or simply the \textit{trajectory}.

The state space \( \mathcal{X} \) contains \( N_{\mathrm{obs}} \) obstacles, denoted by \( \mathcal{X}^{\mathrm{obs}}_i \subset \mathcal{X} \setminus \mathcal{X}_0 \) for \( i = 1, \dots, N_{\mathrm{obs}} \). Each obstacle \( \mathcal{X}^{\mathrm{obs}}_i \) is defined as  
\begin{align}
\mathcal{X}^{\mathrm{obs}}_i = \{ x \in \mathcal{X} : b(x; \xi_i) \leq 0 \}, \quad i \in \mathbb{N}_{1:N_{\mathrm{obs}}},
\end{align}  
where \( b(\cdot; \xi_i): \mathbb{R}^{n_x} \to \mathbb{R} \) is a continuous function parameterized by \( \xi_i \in \mathbb{R}^{n_\xi} \), which encodes the geometric properties of the obstacle region \( \mathcal{X}^{\mathrm{obs}}_i \), such as its position and size. Before control execution, the parameters \( \xi_i \), for \( i \in \mathbb{N}_{1:N_{\mathrm{obs}}} \), are sampled from a probability distribution \( p_{\mathrm{obs}}(\cdot) \), i.e., $\Xi \sim p_{\mathrm{obs}}(\cdot)$, where \( \Xi = [\xi_1, \dots, \xi_{N_{\mathrm{obs}}}] \) represents the collection of sampled parameters defining the \( N_{\mathrm{obs}} \) obstacles. Once generated, these obstacles remain fixed throughout the execution of the control sequence.

\subsection{Specification and optimal control problem}
In this paper, we aim to design a control policy  that generates a trajectory satisfying a given STL specification \(\phi\), i.e., \(x_{0:T} \models \phi\). Given that the obstacles are sampled $\Xi = [\xi_1, ..., \xi_{N_{\mathrm{obs}}}] \sim p_{\mathrm{obs}}(\cdot)$, the specification is defined as follows: 
\begin{align}
   \phi_\Xi = \psi \wedge \left (\bigwedge^{N_{\mathrm{obs}}}_{i=1} \bm{G}_{[0,T]} \neg \mu_{\xi_i} \right ), 
\end{align}
where
\(\psi\) is a given sub-STL formula within \(\phi\), and \(\mu_{\xi_i}\), \(i \in \mathbb{N}_{1: N_{\mathrm{obs}}}\) are the predicates associated with the functions defining the obstacle regions \(b(\cdot; \xi_i)\), \(i \in \mathbb{N}_{1: N_{\mathrm{obs}}}\).

Thus, the objective is to satisfy the STL task $\psi$, while avoiding all obstacles $\mathcal{X}^{\mathrm{obs}}_{i}$ for all $i \in \mathbb{N}_{1: N_{\mathrm{obs}}}$. 
We focus on synthesizing a control policy that generates trajectories satisfying \(\phi\), where the initial state $x_0$ and the obstacle parameters $\xi_i$, \(i \in \mathbb{N}_{1: N_{\mathrm{obs}}}\), are randomly generated according to the distributions \(p_{x_0}(\cdot)\) and \(p_{\mathrm{obs}}(\cdot)\), respectively. 
More specifically, the control policy is derived by solving the following optimal control problem:
\begin{align}
& \mathop{\mathrm{min}}_{W}\ \mathbb{E}_{p_{x_0}, p_{\mathrm{obs}}(\cdot)} \left[ 
 - \rho^{\phi_\Xi} (x_{0:T}) + \gamma \sum_{k=0}^{T-1} g(x_k, u_k) \right] \label{eq:ocp} \\ 
 &\text{subject to:} \notag \\ 
 &\quad u_k = \pi(x_{0:k}; W) \in \mathcal{U},\ \forall k = 0, 1, \dots, T-1\\
&\quad x_{k+1} = f(x_k, u_k),\ \forall k = 0, 1, \dots, T-1, \label{eq:controlconst}
\end{align}
where \(\mathbb{E}_{p_{x_0}, p_{\mathrm{obs}}(\cdot)}\) represents the expected value under the probability distributions \(p_{x_0}\) and \(p_{\mathrm{obs}}\), \(\pi(\cdot; W)\) is the control policy parameterized by \(W\), \(g(\cdot)\) is a continuously differentiable stage cost function, and \(\gamma > 0\) is a given weighting parameter. The objective is to minimize the expected value of the robustness measure \(\rho^{\phi_\Xi}(x_{0:T})\) combined with the control cost, subject to the constraints that the control input is given from \(\pi(\cdot; W)\) and that the trajectory follows the system dynamics. Given that the expectation is taken over  \(p_{x_0}\) and \(p_{\mathrm{obs}}\), the control policy is designed to be \textit{general}, accommodating variability in both initial states and obstacle characteristics. 
As indicated in \eqref{eq:controlconst}, the control policy \(\pi(\cdot; W)\) generates the control input \(u_k\) at time \(k\) based on the \textit{trajectory} \(x_{0:k}\). This is natural since the satisfaction of \(\phi\) depends not only on the current state \(x_k\) but also on the history: \(x_0, \dots, x_{k-1}\).

\begin{figure}[t]
	\centering
	\includegraphics[width=70mm]{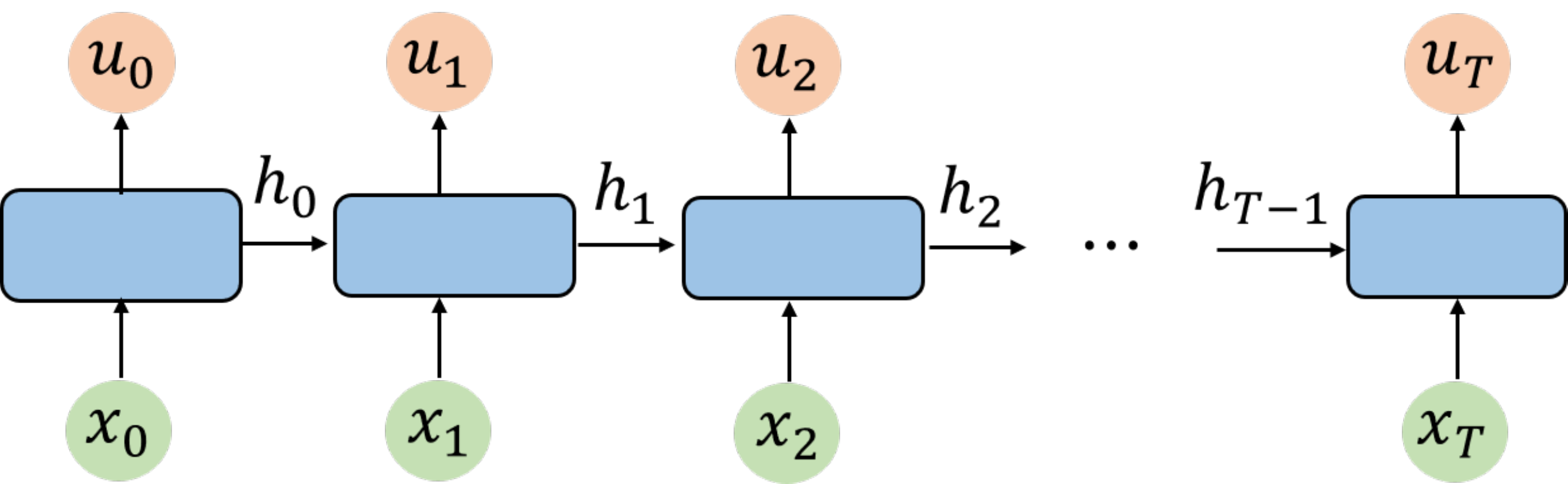}
	\caption{Structure of the RNN.}
	\label{RNN}
\end{figure}

\subsection{Parametrization of the control policy $\pi$}
In this paper, we employ recurrent neural networks (RNNs) for training the control policy $\pi$. RNN is a type of neural network with a recursive structure, characterized by their ability to accept sequential time series data as input. RNNs process data in time series order, retaining information from previously input data while processing later input data, allowing them to address problems with STL constraints where the control actions may vary based on past trajectories. A basic structure of the RNN is illustrated in \rfig{RNN}. 
As shown in \rfig{RNN}, RNN 
keeps processing the sequential information via a hidden state $h_k$, $k\in \mathbb{N}$. 
The update rule of the hidden state and the output (i.e., control input) at time $k$ are as follows: 
\begin{align}
    h_k = l_{W_1} (h_{k-1}, x_k),\ u_k= l_{W_2} (h_{k}), \label{hteq} 
\end{align}
where $l_{W_1}$ and $l_{W_2}$ are the functions parameterized by weights $W_1$ and $W_2$, respectively. Hence, the set of the RNN parameters is given by $W = \{W_1, W_2\}$. The output layer of the function $l_{W_2}$ is characterized as
\begin{gather}
  l_{\mathrm{out}}(x) = \frac{u_{\text{max}} - u_{\text{min}}}{2} \tanh(x) + \frac{u_{\text{max}} + u_{\text{min}}}{2}. 
\end{gather}
Applying the above function allows us to restrict the range of the control inputs from $u_{\text{min}}$ to $u_{\text{max}}$; consequently, it satisfies the control input constraint $u \in [u_{\min}, u_{\max}]$. 

%In this paper, we propose to introduce a \textit{switching logic} so as to improve the control performance based on {clustering} the optimal trajectories. A concrete procedure of the proposed approach is elaborated in the following section. 

\begin{comment}
    
\begin{figure}
\centering
	\subfigure[$\theta = 0^{\circ}$]{%
		\includegraphics[clip, width=0.34\columnwidth]{Figs/optimal1.pdf}}%
	\subfigure[$\theta = 30^{\circ}$]{%
		\includegraphics[clip, width=0.34\columnwidth]{Figs/optimal2.pdf}}%
  \subfigure[$\theta = 60^{\circ}$]{%
		\includegraphics[clip, width=0.34\columnwidth]{Figs/optimal3.pdf}}
	\caption{Examples of optimal trajectories from different initial angles.}
\label{init_angle_and_trajectory}
\end{figure}
\end{comment}

\begin{figure}[t]
\centering

% (a)
\begin{minipage}[b]{0.32\columnwidth}
    \centering
    \includegraphics[width=\linewidth]{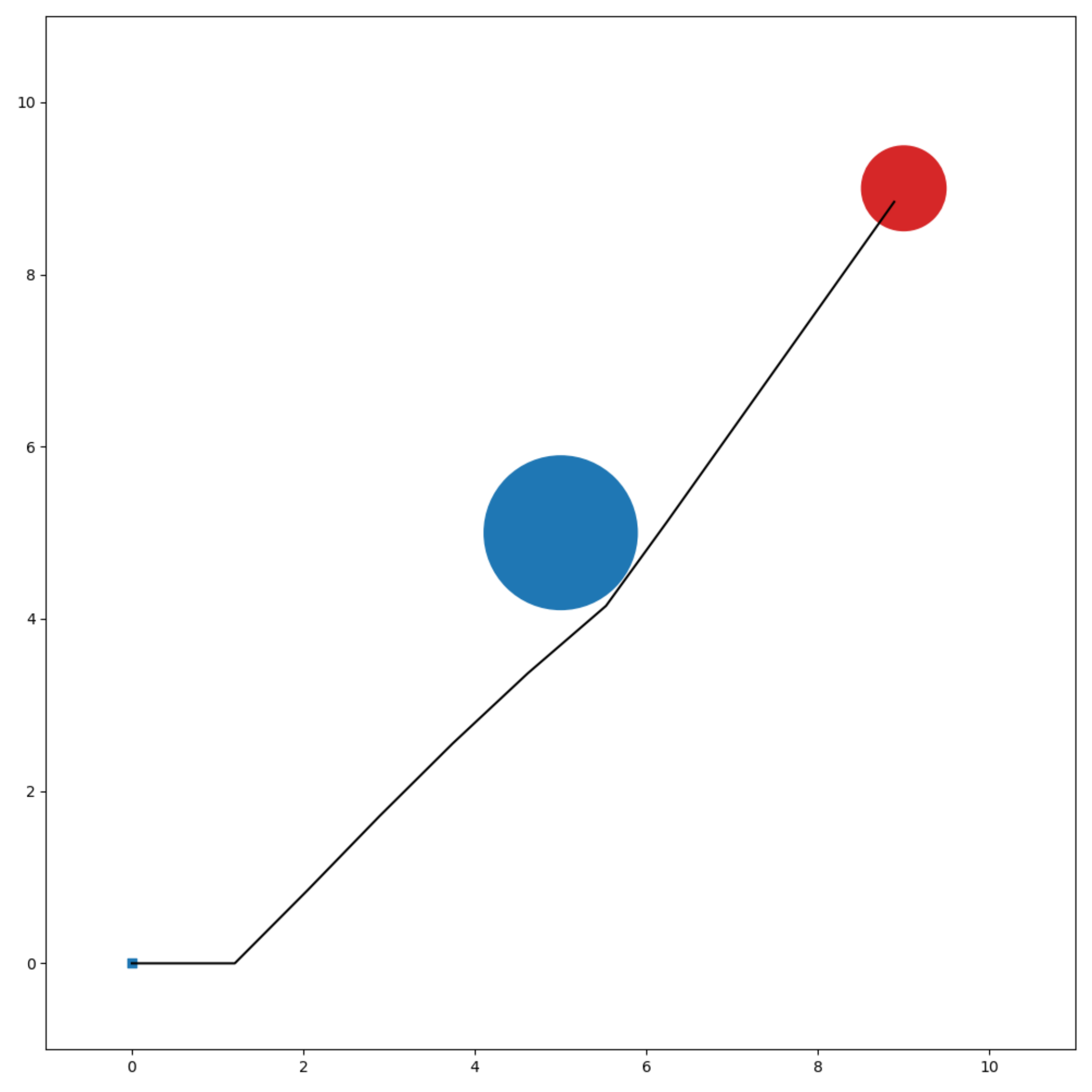}
    \captionof*{figure}{\small (a) $\theta = 0^\circ$}
\end{minipage}
% (b)
\begin{minipage}[b]{0.32\columnwidth}
    \centering
    \includegraphics[width=\linewidth]{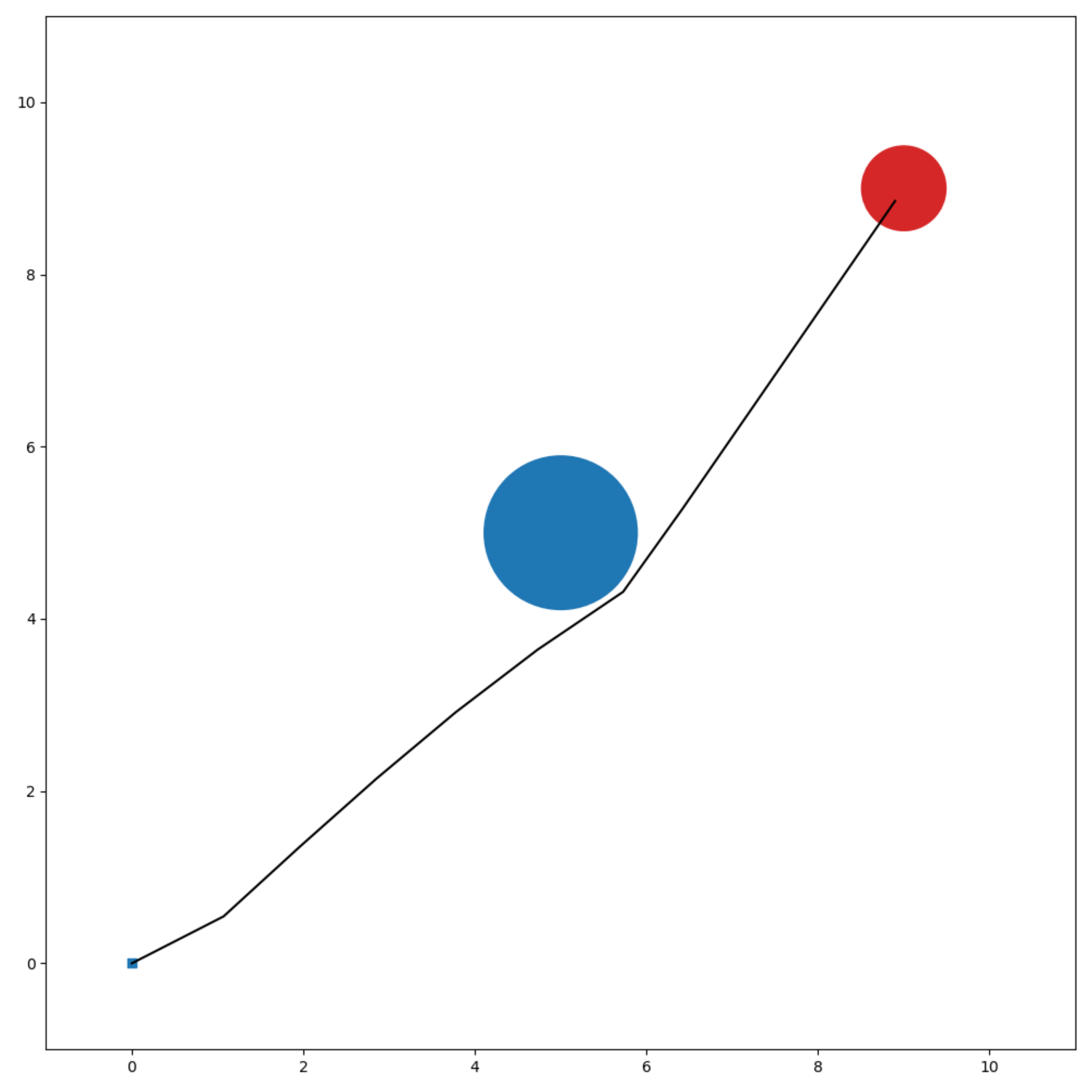}
    \captionof*{figure}{\small (b) $\theta = 30^\circ$}
\end{minipage}
% (c)
\begin{minipage}[b]{0.32\columnwidth}
    \centering
    \includegraphics[width=\linewidth]{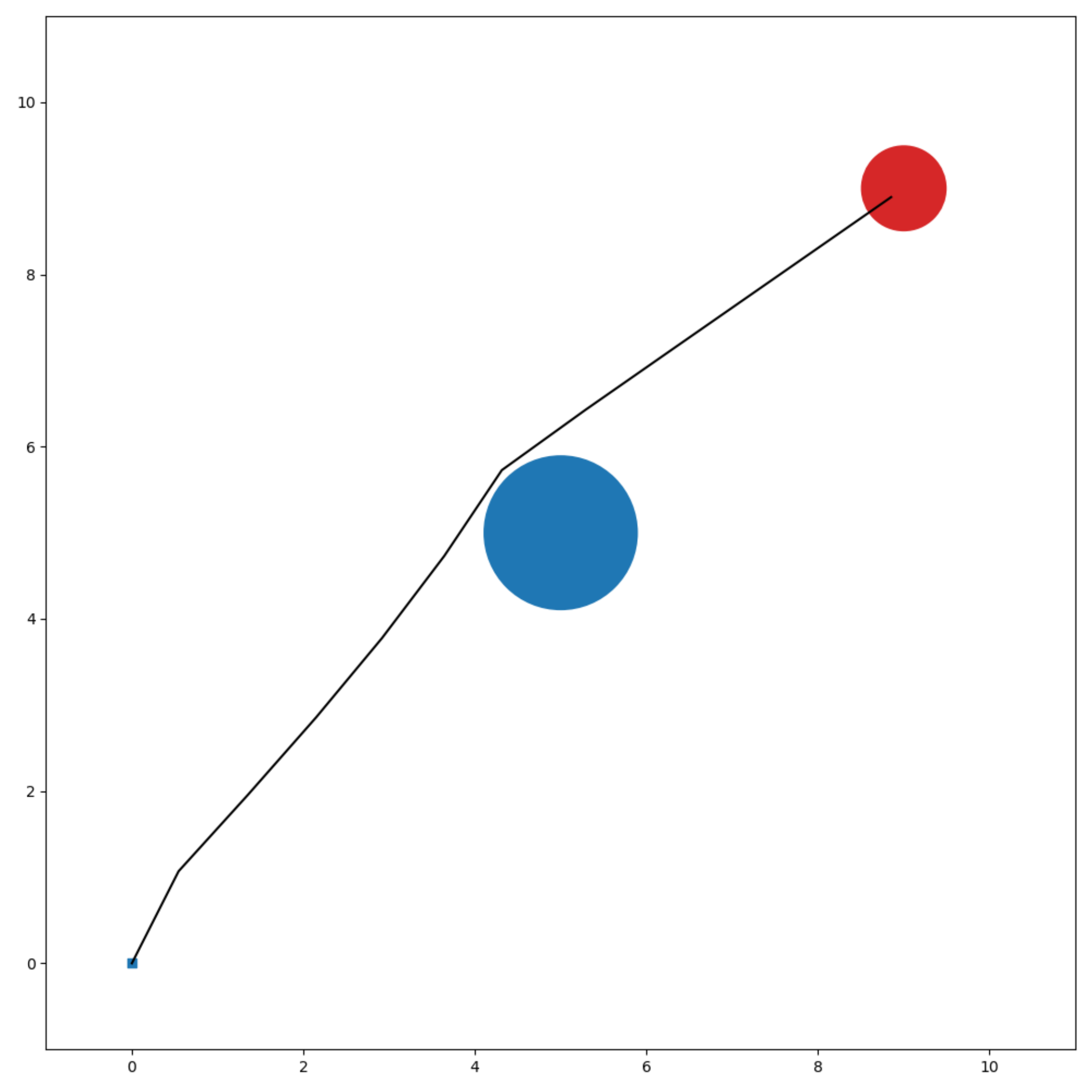}
    \captionof*{figure}{\small (c) $\theta = 60^\circ$}
\end{minipage}

\caption{Examples of optimal trajectories from different initial angles.}
\label{init_angle_and_trajectory}
\end{figure}

\section{Clustering-based RNN controller  synthesis for STL specifications}
In this section, we present a concrete framework for training the RNN control policy  $\pi$ as a solution to the optimal control problem \eqref{eq:ocp}--\eqref{eq:controlconst}. 
\subsection{Motivation}
As presented in \eqref{eq:ocp}--\eqref{eq:controlconst}, we consider an optimal control problem in which the initial state and obstacles are sampled from the probability distributions \(p_{x_0}(\cdot)\) and \(p_{\mathrm{obs}}(\cdot)\), respectively. To illustrate the necessity of \textit{clustering}, we examine a simple reach-avoid task depicted in Fig.~\ref{init_angle_and_trajectory}, where a single obstacle is placed at a fixed location for all initial states to simplify the scenario. We consider three different initial angles: (a) \(\theta = 0^{\circ}\), (b) \(\theta = 30^{\circ}\), and (c) \(\theta = 60^{\circ}\), with the corresponding optimal trajectories shown in Fig.~\ref{init_angle_and_trajectory} (for details on the system dynamics, refer to the numerical simulation in Section~\ref{sec:numerical}). A comparison of trajectories (a) and (b) reveals that both pass to the right of the obstacle, indicating that the change in the initial angle has a minor impact on the trajectory. However, when comparing trajectories (b) and (c), despite the initial angle change being identical to that between (a) and (b), trajectory (c) passes to the left of the obstacle, demonstrating a significant trajectory shift. This observation suggests the existence of a critical initial angle between (b) and (c) at which the trajectory changes \textit{discontinuously}.  
Due to this discontinuity, training a \textit{single} RNN-based control policy to handle all optimal solution trajectories may lead to reduced estimation accuracy and degraded control performance.  

To address this issue, we propose a \textit{clustering} approach for categorizing solution trajectories derived from the optimal control problem based on various initial states and obstacle parameters \(\xi_i\), \(i \in \mathbb{N}_{1:N_{\mathrm{obs}}}\). A dedicated control policy is then assigned to each cluster. Specifically, we train a number of control policies equal to the number of clusters, with each policy trained exclusively on a dataset of similar trajectories. This clustering-based approach enables each control policy to specialize in a specific subset of trajectory data. By switching between these policies based on \((x_0, \Xi)\), we can achieve improved control performance compared to training a single RNN policy.  

\begin{figure}[t]
	\centering
	\includegraphics[width=90mm]{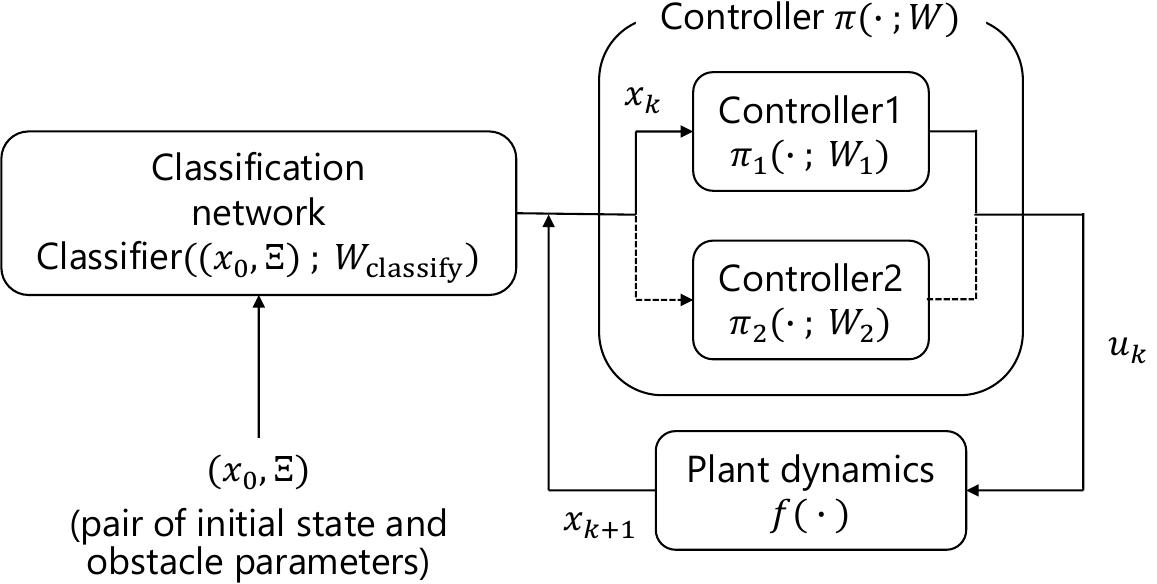}
	\caption{Overview of our approach for the case where the number of clusters is $2$.}
	\label{fig:proposedmethod}
\end{figure}

For instance, if two clusters are identified, the overall feedback control system is structured as shown in Fig.~\ref{fig:proposedmethod}. The details of the \textit{classification network} in Fig.~\ref{fig:proposedmethod} and the mechanism for selecting the appropriate control policy during execution are discussed in the following subsections.  

\subsection{Converting optimal trajectories into feature vectors}
\begin{figure}[t]
	\centering
	\includegraphics[width=70mm]{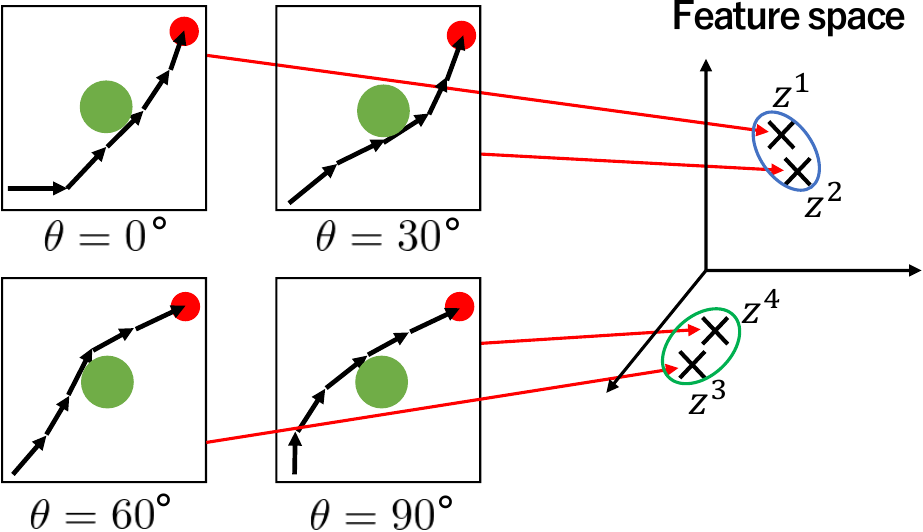}
	\caption{Illustration of transforming optimal trajectories into the feature vectors.}
	\label{fig:transform}
\end{figure}

Since the optimal trajectory obtained from \eqref{eq:ocp}--\eqref{eq:controlconst} is a \textit{time-series} of vectors, standard clustering techniques such as the k-means algorithm, which are applicable only to \textit{a vector}, cannot be directly applied. Therefore, we propose a method to convert the optimal trajectory into a \textit{feature vector} while retaining the necessary information, and then apply clustering to these feature vectors (Fig.~\ref{fig:transform}). 
In this paper, we employ a method that converts trajectories into feature vectors by measuring the \textit{similarity} between pairs of trajectories. This approach is outlined in Algorithm~\ref{embedding}. 
First, we generate $N$ optimal trajectories starting from various initial states and the parameters, i.e., for all $n \in \mathbb{N}_{1:N}$, generate the initial state and the obstacle parameters $x^{n}_0 \sim p_{x_0}(\cdot)$, $\Xi_n \sim p_{\mathrm{obs}}(\cdot)$, and then solve the following optimization problem:
\begin{align}
&\mathop{\mathrm {min}}_{u_0, \ldots, u_{T-1}}\ -\rho^{\phi_{\Xi_n}} \left(x_{0:T}\right)+ \gamma \sum_{k=0}^{T-1} g(x_k, u_k) \notag \\
&\mathrm{s.t.} \ \ x_0 = x^{n}_0, \ {x}_{k+1} = {f}\left({x}_k,u_k\right),\ u_k \in \mathcal{U} \notag \\
&\quad \quad k=0,1,\dots, T-1 \label{eq:optimization}. 
\end{align}

Let \(x_{0:T}^{n}= x_0^{n}, x_1^{n}, \dots, x_T^{n}\) denote the  \(n\)-th  optimal trajectory obtained by solving the optimization problem \eqref{eq:optimization} 
 from the \(n\)-th initial states $x^{n}_0$ and obstacle parameters $\Xi_n$.

Given this set of  $N$ optimal trajectories generated from various initial states and obstacle parameters, we define the \textit{similarity} \(D(n, m)\) between the \(n\)-th and \(m\)-th optimal trajectories (\(n, m \in \{1, \dots, N\}\)),  obtained from the corresponding initial states and obstacle parameters,  as the cumulative error between them:
\[
D(n, m) = \sum_{k = 0}^{T} \|x_{k}^{n} - x_{k}^{m}\|_M,
\]
where \(M \in \mathbb{R}^{n_x \times n_x}\) is a positive semi-definite weighting matrix. The feature vector \(z_n\), derived from the trajectory \(x_{0:T}^{n}\), is then defined as the collection of similarities between the \(n\)-th trajectory and all other trajectories in the dataset:
\[
z^n = [D(n,1), D(n,2), \dots, D(n,N)]^\top \in \mathbb{R}^N.
\]
Standard clustering techniques, such as the k-means \cite{kmeansalgorithm}, X-means algorithm \cite{xmeansalgorithm}, are subsequently applied to the feature vectors \(z^n\), for \(n \in \mathbb{N}_{1:N}\), in order to group similar trajectories into clusters.

As an example, if the \(n_1\)-th trajectory \(x_{0:T}^{n_1}\) and the \(n_2\)-th trajectory \(x_{0:T}^{n_2}\) are similar, i.e., \(x_{0:T}^{n_1} \approx x_{0:T}^{n_2}\), then for any \(i \in \mathbb{N}_{1:N}\), the corresponding similarity measures \(D(n_1, i)\) and \(D(n_2, i)\) will also be close in value, i.e., \(D(n_1, i) \approx D(n_2, i)\). In this case, the feature vectors \(z_{n_1}\) and \(z_{n_2}\), defined as
\begin{align}
&z^{n_1} = [D(n_1,1), D(n_1,2), \dots, D(n_1,N)]^\top, \\ 
&z^{n_2} = [D(n_2,1), D(n_2,2), \dots, D(n_2,N)]^\top,
\end{align}
will have components that are similar across all indices, making it likely that these trajectories will be grouped into the same cluster during the clustering process.

\begin{algorithm}[t]
  \caption{Feature vector encodings of optimal trajectories based on similarities}
  \label{embedding}
 % \textbf{Input} : $x^i _k, k=0,...,T \; (i = 1, \dots, N)$ \\
  \textbf{Output} : $z_n \; (n = 1, \dots, N)$
  \begin{algorithmic}[1]
    %\STATE \textbf{Initialize} : $x^i _k \; (i = 1, 2, \dots, N, \; k = 0, 1, \dots, T)$
    \FOR{$n =1 : N$}
    \STATE Sample $x^{n}_0 \sim p_{x_0}(\cdot)$, $\Xi_n \sim p_{\mathrm{obs}}(\cdot)$
    \STATE Solve \eqref{eq:optimization} to get the optimal trajectory $x^n_0, ..., x^n_{T-1}$
    \ENDFOR
    \STATE $n \leftarrow 1$
    \WHILE{$n \leq N$}
      \STATE $m \leftarrow 1$
      \WHILE{$m \leq N$}
        \STATE $D(n, m) \leftarrow  \sum_{k = 0}^{T} \|x^n _k - x^m _k \|_{M}$
        \STATE $m \leftarrow m + 1$
      \ENDWHILE
      \STATE $n \leftarrow n + 1$
    \ENDWHILE
    
    \STATE $n \leftarrow 1$
    \WHILE{$n \leq N$}
      \STATE $z_n \leftarrow \left[ D(n, 1),\ D(n, 2),\  \cdots ,\  D(n, N) \right]^\top$
      \STATE $n \leftarrow n + 1$
    \ENDWHILE
  \end{algorithmic}
\end{algorithm}

\subsection{Classification network}
\subsubsection{Motivation and structure}
As discussed in the previous section, the data used for clustering consists of a set of feature vectors derived from optimal trajectories corresponding to various sampled initial states and obstacle parameters. For pairs of initial states and obstacle parameters that have already been sampled and included in the clustering process, i.e., \((x^n_0, \Xi_n )\) for \(n = 1, 2, \ldots, N\), the assigned clustering labels are readily available. However, for {unobserved} (un-sampled) pairs, i.e., \((x_0, \Xi) \neq (x^n_0, \Xi_n)\) for all \(n = 1, 2, \ldots, N\), the clustering label cannot be directly determined. To address this issue, we propose a neural network-based approach to classify the appropriate clustering label for \textit{any} given pair \((x_0, \Xi)\). This network, referred to as the \textit{classification network}, is designed with an input dimension of \(n_x + n_{\xi} N_{\mathrm{obs}}\) and an output dimension equal to the number of clustering labels, denoted as \(n_c\). The input-output relationship of the classification network is defined as follows:  
\begin{align}\label{eq:ell}
\ell = \text{Classifier}((x_0, \Xi); W_{\text{classify}}) := \underset{\ell \in \{1,...,n_c\}}{\mathrm{argmax}} \{{y}_\ell \},
\end{align}
where \({y}_\ell\) for \(\ell \in \{1,...,n_c\}\) is computed as  
\[
[{y}_1, ..., {y}_{n_c}]^\top = \text{softmax}\left( \mathcal{N\!N}\left((x_0, \Xi); W_{\text{classify}} \right) \right),
\]
with \(\text{softmax}(a) = \frac{\exp(a)}{\sum \exp(a)}\). Here, \(\ell\) represents the predicted clustering label for the input \((x_0, \Xi)\), while \(\mathcal{N\!N}(\cdot;W_{\text{classify}} )\) denotes the neural network parameterized by \(W_{\text{classify}}\).  

\begin{figure}[t]
	\centering
\includegraphics[width=80mm]{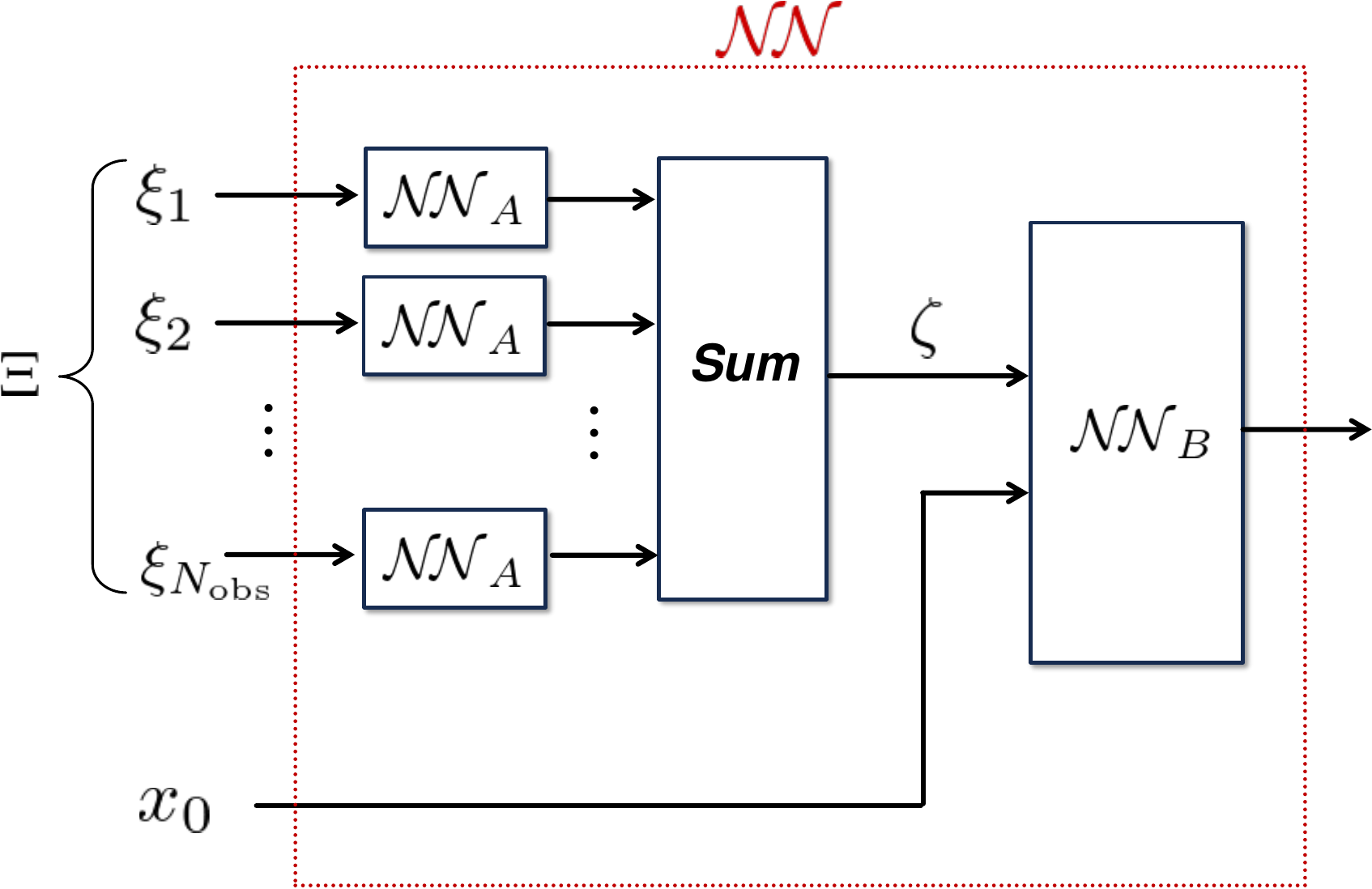}
	\caption{Illustration of $\mathcal{N\!N}$ preserving permutation invariance with respect to $\Xi$.}
	\label{fig:NN_architecture}
\end{figure}

Since the geometric properties of the environment remain \textit{invariant} under different orderings of the obstacle parameters \(\Xi\), the classification network must preserve this permutation invariance. For instance, consider \(\xi_1 = [1, 2]^\top\), \(\xi_2 = [3, 4]^\top\), and their reordered counterparts \(\xi'_1 = [3, 4]^\top\), \(\xi'_2 = [1, 2]^\top\). Transitioning from \(\Xi = [\xi_1, \xi_2]\) to \(\Xi' = [\xi'_1, \xi'_2]\) does not alter the geometric properties of the environment. To enforce this invariance, we introduce an additional layer that ensures \textit{permutation invariance} in the obstacle parameter \(\Xi\). Specifically, following \cite{NIPS2017deepsets}, we incorporate this property into the classification network using the following formulation:
\begin{align} 
\mathcal{N\!N}\left((x_0, \Xi); W_{\text{classify}} \right) = {\mathcal{N\!N}}_B\left((x_0, \zeta); W_{\text{classify}} \right), 
\end{align}
where \(\zeta\) is defined as
\begin{align}
\zeta =  \sum_{j} \mathcal{NN}_A (\xi_j ;W_{\text{classify}} ),
\end{align}
with \(\Xi = [\xi_1,...,\xi_{N_{\mathrm{obs}}}]\). \(\mathcal{NN}_A\) and \(\mathcal{NN}_B\) denote neural networks responsible for processing obstacle parameters and classification, respectively. The architecture of \(\mathcal{N\!N}\) is illustrated in Fig.~\ref{fig:NN_architecture}. This structure ensures that the network preserves permutation invariance with respect to the environmental parameter $\Xi$. 
\subsubsection{Training the classification network}
To learn the weight parameter \(W_{\text{classify}}\), we first construct the training dataset. For each pair \((x^n_0, \Xi_n)\), the optimal trajectories obtained from \eqref{eq:optimization} are converted into feature vectors, and the resulting clustering labels are denoted as \(\ell_n \in \{1, 2, \ldots, n_c\}\). Each label \(\ell_n\) is then represented as a \textit{one-hot vector} \(y^n \in \{0, 1\}^{n_c}\), where the \(\ell_n\)-th element is 1 and all other elements are 0. The training dataset consists of input-output pairs \(\{ (x^n_0, \Xi_n), y^n\}\), for \(n = 1, \ldots, N\). The training procedure for \(W_{\text{classify}}\) is summarized in Algorithm~2.  

During training, we first compute the classification loss \(\mathcal{L}_{\text{classify}}\) using the current parameter \(W_{\text{classify}}\), where the cross-entropy loss is employed (line 3, Algorithm~2):  
\begin{gather}
  \mathcal{L}_{\text{classify}} = - \frac{1}{N} \sum_{n=1}^{N} \sum_{\ell =1}^{n_c} y^n _{\ell} \ln \hat{y}^n _{\ell}, \label{eq:loss_classifier}
\end{gather}
where \(y^n _{\ell}\) denotes the \(\ell\)-th element of \(y^n\) and  
\begin{align}
    \hat{y}^n = 
  \begin{bmatrix}
    \hat{y}^{n} _1\\
    \hat{y}^{n}_2\\
    \vdots\\
    \hat{y}^{n}_{n_c}\\
  \end{bmatrix}
  = \text{softmax}\left( \mathcal{N\!N}\left( (x^n_0, \Xi_n); W_{\text{classify}} \right) \right). \label{y^i}
\end{align}
Subsequently, the parameter \(W_{\text{classify}}\) is updated using the gradient descent (backpropagation) to minimize the cross-entropy loss \eqref{eq:loss_classifier} (line 5, Algorithm~2). By iteratively evaluating the loss and updating the parameters over \(N_{\mathrm{epoch}}\) training iterations, the classification network is optimized.

\begin{algorithm}[t]
  \caption{Learning the classification network}
  \label{classifier}
  \textbf{Input} : $\{ (x_{0}^{n}, \Xi_n), y^n\}, \;  n = 1, 2, \dots, N$ (training data for classification),\  $N_{\mathrm{epoch}}$ (Number of epochs), $\alpha >0$ (parameter for the gradient decent)\\
  \textbf{Output} : $W_{\text{classify}}$ (weight parameter for classification network)
  \begin{algorithmic}[1]
  \STATE Initialize $W_{\text{classify}}$;
\STATE $\text{epoch} \leftarrow 1$
    \WHILE{$\text{epoch} \leq N_{\mathrm{epoch}}$}
      \STATE Compute 
 $\mathcal{L}_{\text{classify}}$ in \eqref{eq:loss_classifier}
      \STATE Update the weight parameter as $W_{\text{classify}} \leftarrow W_{\text{classify}} - \alpha \partial L_{\text{classify}}/\partial W_{\text{classify}}$
      \STATE $\text{epoch} \leftarrow \text{epoch} + 1$
    \ENDWHILE
  \end{algorithmic}
\end{algorithm}

\subsection{Training RNN control policies}
Now we proceed to train the RNN control policies. During training, each RNN control policy is assigned to a specific cluster identified in the previous subsections. Specifically, for each \(\ell = 1, 2, \dots, n_c\), we denote the RNN control policy assigned to clustering label \(\ell \in \{1, ..., n_c\}\) as \(\pi_\ell (\cdot; W_\ell)\), where \(W_\ell\) denotes the weight for \(\pi_\ell\).

During control execution, the appropriate control policy from the set of candidates \(\pi_\ell\), \(\ell \in \{1, ..., n_c\}\), is selected based on the predicted label obtained from the classification network. Specifically, given any \(x_0 \sim p_{x_0}(\cdot)\) and \(\Xi = [\xi_1, ..., \xi_{N_{\mathrm{obs}}}] \sim p_{\mathrm{obs}}(\cdot)\) and letting \(\ell = \text{Classifier}((x_0, \Xi); W_{\text{classify}})\), the control policy is defined as follows:
\begin{eqnarray}
  \pi(x; W) = \begin{cases}\label{equ:control}
    \pi_1(x; W_1)\ \quad (\text{if}\ \ell=1) \\
    \pi_2(x; W_2)\ \quad (\text{if}\ \ell=2) \\
    \quad \quad \quad \quad  \quad \quad \quad\vdots \\
    \pi_{n_c}(x; W_{n_c})\ (\text{if}\ \ell=n_c), \\
  \end{cases}
\end{eqnarray}
where \(W = \{W_1, ..., W_{n_c}\}\).

To prepare for training the RNN control policy  \(\pi_\ell\), we divide the training data into \(n_c\) subsets based on the classification network. This procedure is detailed in Algorithm~\ref{dataset}. First, we generate \(N\) pairs of initial states and obstacle parameters \((x^n_0, \Xi_n),\ n = 1, 2, \ldots, N\), and then obtain the corresponding predicted labels using the classification network \(\ell_n,\ n = 1, 2, \ldots, N\) (lines 2–5). Next, we create a new dataset based on the predicted labels as follows (lines 7–18):
\begin{align}
    &\text{Cluster 1}:\ \  (x_{0,1}^j, \Xi_{j,1}) \quad \ \ (j = 1,2, \ldots , N_1) \notag \\
    &\text{Cluster 2}:\ \ (x_{0,2}^j, \Xi_{j,2}) \quad \ \ (j = 1,2, \ldots , N_2) \notag  \\
    &\quad \quad \quad \quad \quad \quad \quad \quad \vdots \label{equ:class_data} \\
   & \text{Cluster \(n_c\)}: (x_{0,n_c}^j, \Xi_{j,n_c}) \quad (j = 1,2, \ldots , N_{n_c}), \notag 
\end{align}
where \(N_\ell\) denotes the number of training data points assigned for the clustering label \(\ell\).
In \eqref{equ:class_data}, the data in each row \((x_{0,\ell}^j, \Xi_{j,\ell}), \ j = 1,2, \ldots, N_{\ell}\) is used for training the RNN control policy  \(\pi_\ell (\cdot; W_\ell)\).
%In other words, we train the RNN control policy  $\pi_\ell(\cdot; W_{\ell})$, $\ell = 1, ..., N_{n_c}$, where $W_{\ell}$ is the NN parameter to characterize $\pi_\ell$,and$x_{0,\ell}^{i} \; (i = 1, 2, \dots, N_\ell)$ is assigned as the training data to learn $\pi_\ell$. 

\begin{algorithm}[t]
  \caption{Dataset construction for RNN control policies}
  \label{dataset}
  \textbf{Output} : $(x_{0,\ell}^{j}$,$\Xi_{j,\ell})$, $j = 1, \dots, N_\ell$, $\ell=1,...,N_\ell$ (initial dataset to train the RNN control policy $\pi_\ell$, $\ell=1,...,N_\ell$)
  \begin{algorithmic}[1]
    %\STATE \textbf{Initialize} : $(x_{0}^{i}, P^i) \; (i = 1, 2, \dots, N), \; \text{Classifier}(\cdot; W_{\text{classify}}), \;  n$
    
    \STATE $n \leftarrow 1$
    \WHILE{$n \leq N$}
      \STATE $x_{0}^{n} \sim p_{x_0}(\cdot)$, $\Xi_n \sim p_{\mathrm{obs}}(\cdot)$
      \STATE $\ell_n = \text{Classifier}((x_{0}^{n},\Xi_n); W_{\text{classify}})$
    \ENDWHILE
    \STATE $\ell \leftarrow 1$
    
    \WHILE{$\ell \leq n_c$}
      \STATE $i \leftarrow 1, \; j \leftarrow 1$
      \WHILE{$i \leq N$}
        \IF{$\ell_i == \ell$}
          \STATE $(x_{0,\ell}^{j}, \Xi_{j,\ell}) \leftarrow (x_{0}^{i}, \Xi_i)$
          \STATE $j \leftarrow j + 1$
        \ENDIF
        \STATE $i \leftarrow i + 1$
      \ENDWHILE
      \STATE $N_\ell \leftarrow j -1$
      \STATE $\ell \leftarrow \ell + 1$
    \ENDWHILE
  \end{algorithmic}
\end{algorithm}

Using the dataset obtained above, let us now proceed to train the control policies. The training method for the RNN control policy \(\pi_\ell (\cdot; W_\ell)\) is outlined in Algorithm~\ref{controlpolicy}. 
In the algorithm, we iterate the following procedures to optimize the parameter $W_\ell$:  
\begin{itemize}
    \item \textit{Forward computation (lines~4--14):} For fixed $W_\ell$, generate the trajectories $x_{0:T}^{j}$ by iteratively applying  ${x}_{k+1}^{j}={f}({x}_k^{j}, {\pi}_\ell  (x^{j}_{0:t};W_\ell ))$ for all $t=0, ..., T-1$ from the initial state $x_{0}^{j} = x_{0,\ell}^{j}$. Then, we compute the following loss:
    \begin{align}
  & \mathcal{L}_{\mathrm{total}, \ell} = -\mathcal{L}_{\text{robustness}, \ell} + \gamma \mathcal{L}_{\text{control}, \ell} \label{loss_controlpolicy}\\
  &\mathcal{L}_{\text{robustness}, \ell} = \frac{1}{N_\ell} \sum_{j=1}^{N_\ell} \rho^{\phi_{\Xi_j}} \left( x_{0:T}^{j} \right), \\
  &\mathcal{L}_{\text{control}, \ell} = \frac{1}{N_\ell} \sum_{j=1}^{N_\ell} \sum_{k=0}^{T-1} g(x_k^j, u_k^j), \label{eq:L_control}
\end{align}
where $\Xi_j = \Xi_{j,\ell}$. 
The computed total loss $\mathcal{L}_{\mathrm{total}, \ell}$ represents an approximation of the expectation in \eqref{eq:ocp}. 
    \item \textit{Backward computation and parameter update (lines~15,16):} Compute the gradients for all the parameters via backpropagation through time (BPTT) \cite{LSTM}. This computation can be easily implemented, e.g., by combining the auto-differentiation tools designed for NNs (e.g., \cite{paszke2017automatic}) and STLCG \cite{leung2023backpropagation}. 
    Thereafter, we update all the parameters $W_\ell$ by through several well-known techniques, e.g., Adam optimizer \cite{kingma2014adam}.
\end{itemize}

\begin{algorithm}[t]
  \caption{Algorithm to train the control policy $\pi_\ell$}
  \label{controlpolicy}
  \textbf{Input} : $(x_{0,\ell}^{j},\Xi_{j,\ell})\; 
  (j = 1, 2, \dots, N_\ell)$, $N_{\mathrm{epoch}}$ (number of epochs), $\beta >0$ (parameter for the gradient decent)\\
  \textbf{Output} : $W_\ell$ (weight for $\pi_\ell$)
  \begin{algorithmic}[1]
    %\STATE \textbf{Initialize} : $(x_{0,c}^{i}, P_{c}^{i}) \; (i = 1, 2, \dots, N_c), \; \text{num\_epochs}, \; T$
    
    \STATE $\text{epoch} \leftarrow 1$
    \WHILE{$\text{epoch} \leq N_{\mathrm{epoch}}$}
      \STATE $j \leftarrow 1$\\
\nonumber{\textbf{[Forward computation]:}}
      %\STATE $L_{\mathrm{total}} \leftarrow 0$
      \WHILE{$j \leq N_\ell$}
            \STATE $x_{0}^{j} = x_{0,\ell}^{j}$, $\Xi_j = \Xi_{j,\ell}$
        \STATE $k \leftarrow 0$
        \WHILE{$k \leq T - 1$}
          \STATE $u_{k}^{j} = \pi_\ell (x_{0:k}^{j}; W_{\ell})$
          \STATE $x_{k+1}^{j} = f(x_{k}^{j}, u_{k}^{j})$
          \STATE $k \leftarrow k + 1$
        \ENDWHILE
        \STATE $j \leftarrow j + 1$
      \ENDWHILE
      
      \STATE Compute the loss 
$\mathcal{L}_{\mathrm{total}, \ell}$ according to \eqref{loss_controlpolicy}--\eqref{eq:L_control}\\
\smallskip
\nonumber{\textbf{[Backward computation and parameter update]:}} 
\STATE Through the backpropagation,
compute the gradients with respect to the weight parameters: $\mathcal{L}_{\mathrm{total}}/\partial W_{\ell}$ 
    \STATE Update the parameter as $W_{\ell} \leftarrow W_{\ell} - \beta \partial \mathcal{L}_{\mathrm{total}}/\partial W_{\ell}$
      \STATE $\text{epoch} \leftarrow \text{epoch} + 1$
    \ENDWHILE
  \end{algorithmic}
\end{algorithm}
%However, since this is true for all indices, we only describe it for the control policy  $\pi_c$. 
%preventing the control policy  from outputting inappropriate values as control inputs.

%%%%%%%%%%%%%%%%%%%%%%%%%%%%%%%%%%%%%%%%%%%%%%%%%%%%%%%%%
%%%%%%%%%%%%%%%%%%%%%%%%%%%%%%%%%%%%%%%%%%%%%%%%%%%%%%%%%
%%%%%%%%%%%%%%%%%%%%%%数値実験ここから%%%%%%%%%%%%%%%%%%%%%%
%%%%%%%%%%%%%%%%%%%%%%%%%%%%%%%%%%%%%%%%%%%%%%%%%%%%%%%%%
%%%%%%%%%%%%%%%%%%%%%%%%%%%%%%%%%%%%%%%%%%%%%%%%%%%%%%%%%

\begin{figure}[t]
	\centering
	\includegraphics[width=80mm]{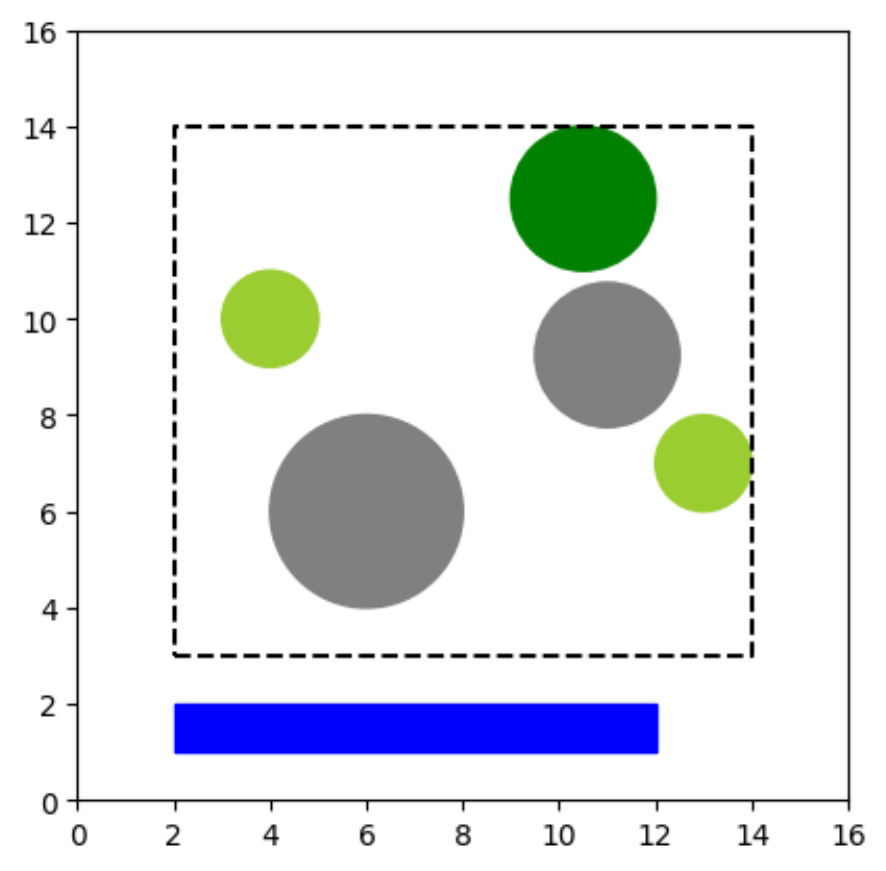}
	\caption{Environment in the numerical experiment. The blue, gray, green, yellow-green regions represent the set of initial positions in $\mathcal{X}_0$, obstacles, the goal region, and the transit regions, respectively.}
	\label{kankyou}
\end{figure}

\section{Numerical Experiment} \label{sec:numerical}
This section presents numerical experiments to evaluate the effectiveness of the proposed method. The performance of the proposed approach is compared with that of conventional methods that do not employ the clustering.
\subsection{Problem Setup}
We consider a path planning problem for a vehicle operating in a two-dimensional (2D) workspace. The vehicle dynamics are modeled as a nonlinear, discrete-time, nonholonomic system:  
\begin{align}
  x_{k+1} = x_k +
  \begin{bmatrix}
    v_k \cos \theta_k\\
    v_k \sin \theta_k\\
    \omega_k\\
    F_k / m\\
    \tau_k / I
  \end{bmatrix}
\end{align}
where the state and control input vectors are defined as $x_k = [p_{1,k}, p_{2,k}, \theta_k, v_k, \omega_k]^\top$ and $u_k = [F_k, \tau_k]^\top$, respectively. The physical meanings of these variables are summarized in Table~\ref{mozi}. 
The environment is illustrated in Fig.~\ref{kankyou}. The stage cost function in \eqref{eq:optimization} is given by $g(x_k, u_k) =\| u_k \|_R$ with $R = [10, 0; 0, 1]$.

The environment consists of three static regions: the initial state region (blue), the goal state region (green), and the transit regions (yellow-green) that serve as intermediate waypoints. As will be detailed later, the task requires the vehicle to navigate from the initial region to the goal region while passing through at least one of the two designated transit (yellow-green) regions. The set of initial states is given by $\mathcal{X}_{\text{init}} = \{[p^{\text{init}}_{x},p^{\text{init}}_y]^\top \in \mathcal{P}_{\text{init}}, \theta \in [-\pi, \pi], v \in [0.5,1], \omega =0 \}$, where $\mathcal{P}_{\text{init}} = [2, 12] \times [1, 2] \subset \mathbb{R}^2$ corresponds to the blue region in Fig.~\ref{kankyou}.  

The obstacle regions (gray) are introduced in the environment and are randomly generated as circular regions:
\begin{align}
    \mathcal{X}^{\text{obs}}_i = \{[p^{\text{obs}}_{x,i}, p^{\text{obs}}_{y,i}]^\top \in \mathcal{P}_{\text{obs}},\  r^{\text{obs}}_i \in [1.5, 2] \},
\end{align}
 for all $i \in \mathbb{N}_{1:{N_{\text{obs}}}}$, where $\mathcal{P}_{\text{obs}}=[2,14]\times[3,14] \subset \mathbb{R}^2$ (depicted as the region enclosed by dotted lines in Fig.~\ref{kankyou}), and $r^{\text{obs}}_i$ represents the radius of the obstacle regions. We assume $N_{\text{obs}}=2$, and the center positions and radius of these obstacles are randomly assigned with the constraint that all obstacles do not overlap with other obstacles or transit regions. In other words, the environmental parameter is defined as $\Xi = [\xi_1, \xi_2]$, where $\xi_i = [p^{\text{obs}}_{x,i}, p^{\text{obs}}_{y,i}, r^{\text{obs}}_i]^\top$ for $i=1,2$. 

The STL task is expressed as  
\begin{align}
    \bigvee^2_{i=1}	F_{[0,T]} \phi_{\text{tran},i} \wedge F_{[0, T]} \phi_{\text{goal}} \wedge \bigwedge^{2}_{i=1} G_{[0, T]} \neg \phi_{\text{obs},i}
    \label{STLtask}
\end{align}
with $T=25$. Here,
$\vee^2_{i=1}F_{[0,T]} \phi_{\text{tran},i} \wedge F_{[0, T]} \phi_{\text{goal}}$ 
represents the STL task of reaching the goal region (green region in Fig.~\ref{kankyou}) within $T$ steps while passing through at least one of the transit regions (yellow-green regions in Fig.~\ref{kankyou}). Additionally, $G_{[0, T]} \neg \phi_{\text{obs},i}$ $(i=1,2)$ and $G_{[0, T]} \neg \phi_{\text{obs2}}$ ensures that the obstacle regions (gray regions in Fig.~\ref{kankyou}) are avoided throughout the planning horizon.

\begin{table}[t]
  \centering
  \caption{Definition of the state, control input and the parameters.}
  \begin{tabular}{ccc}
        \hline
    Symbol & Meaning \\
       \hline \hline
    $p_k = [p_{1,k}, p_{2,k}]^\top$ & Two-dimensional position\\
    $\theta_k$ & Angle  \\
    $v_k$ & Speed \\
    $\omega_k$ & Angular velocity \\
    $F_k$ & Force \\
    $\tau_k$ & Torque \\
    $m=10$ & Mass (constant parameter) \\
    $I=100$ & Moment of inertia (constant parameter)\\
    \hline
  \end{tabular}
  \label{mozi}
\end{table}

\subsection{Some details on clustering, optimization, and neural networks}

For clustering, we employ the X-means method~\cite{xmeansalgorithm}, an extension of the K-means algorithm that automatically determines the optimal number of clusters. 
Unlike K-means, which requires a predefined number of clusters, X-means iteratively adjusts the number based on cluster quality, evaluated using the Bayesian Information Criterion (BIC) (for details, see~\cite{xmeansalgorithm}) or Akaike Information Criterion (AIC). In this study, we use the PyClustering library \cite{pyclustering}, which supports both AIC and BIC as evaluation criteria.
The detailed formulation of the evaluation criterion used in this study is provided in Appendix~\ref{appendix:clustering}.

For the optimization component, we utilize IPOPT within the CasADi framework~\cite{andersson2019casadi} to numerically solve the optimization problem formulated in~\eqref{eq:optimization}. Both the classification network and the RNN controller are implemented using the PyTorch framework \cite{paszke2017automatic}. 
The hyperparameters of the neural network, including the number of hidden units, layers, and training epochs, are summarized in Tables~\ref{table:classification} and \ref{table:RNNcontroller}.

\begin{table}[t]
	\centering
	\caption{Hyper parameters of the classification network. \label{table:classification}}
	\begin{tabular}{l|l}
		\hline
		\(\mathcal{NN}_A\)  hidden state	&  128	\\
		\(\mathcal{NN}_A\)  hidden layer			&  2	\\
            \(\mathcal{NN}_B\)  hidden state	&  64	\\
		\(\mathcal{NN}_B\)  hidden layer			&  2	\\
		learning rate  	&  0.001	\\
		epoch     		&  500	\\
		batch size  	&  32 \\
		\hline
	\end{tabular}
\end{table}

\begin{table}[t]
	\centering
	\caption{Hyper parameters of the RNN controller. \label{table:RNNcontroller}}
	\begin{tabular}{l|l}
		\hline
		hidden state	&  32	\\
            hidden layer    &  1	\\
		learning rate  	&  0.01	\\
		epoch     		&  30	\\
		batch size  	&  8 \\
		\hline
	\end{tabular}
\end{table}

\begin{table}[t]

  \centering
  \caption{Training time of the single RNN approach and the clustering-based RNN approach }
  \begin{tabular}{ccc}
    \hline
    Controller & Training time [s] \\ 
    \hline \hline
    Clustering-based RNN 1 & 2493 \\
    Clustering-based RNN 2 & 3002 \\
    Clustering-based RNN 3 & 3055 \\
    Clustering-based RNN 4 & 2525 \\
        \hline
    Clustering-based RNN total &  11775\\
        \hline
    Single RNN & 12394 \\
    \hline
  \end{tabular}
  \label{time}
\end{table}

\begin{comment}
    
\begin{figure*}[h]
\centering
	\subfigure[Cluster 1]{%
		\includegraphics[clip, width=0.5\columnwidth]{Figs/Optimal/0-575.pdf}}%
	\subfigure[Cluster 1]{%
		\includegraphics[clip, width=0.5\columnwidth]{Figs/Optimal/0-607.pdf}}%
  \subfigure[Cluster 2]{%
		\includegraphics[clip, width=0.5\columnwidth]{Figs/Optimal/1-149.pdf}}%
	\subfigure[Cluster 2]{%
		\includegraphics[clip, width=0.5\columnwidth]{Figs/Optimal/1-328.pdf}}%
\par
  \subfigure[Cluster 3]{%
		\includegraphics[clip, width=0.5\columnwidth]{Figs/Optimal/2-200.pdf}}
 \subfigure[Cluster 3]{%
		\includegraphics[clip, width=0.5\columnwidth]{Figs/Optimal/2-63.pdf}}
  \subfigure[Cluster 4]{%
		\includegraphics[clip, width=0.5\columnwidth]{Figs/Optimal/3-618.pdf}}
  \subfigure[Cluster 4]{%
		\includegraphics[clip, width=0.5\columnwidth]{Figs/Optimal/3-627.pdf}}
	\caption{Optimal trajectories obtained by solving the optimization problem, classified into Cluster 1, Cluster 2, Cluster 3, and Cluster 4 for (a)-(b), (c)-(d), (e)-(f), and (g)-(h), respectively.}
\label{optimal trajectory}
\end{figure*}
\end{comment}

\begin{figure*}[ht]
\centering

% --- 1段目 ---
\begin{minipage}[b]{0.24\textwidth}
    \centering
    \includegraphics[width=\linewidth]{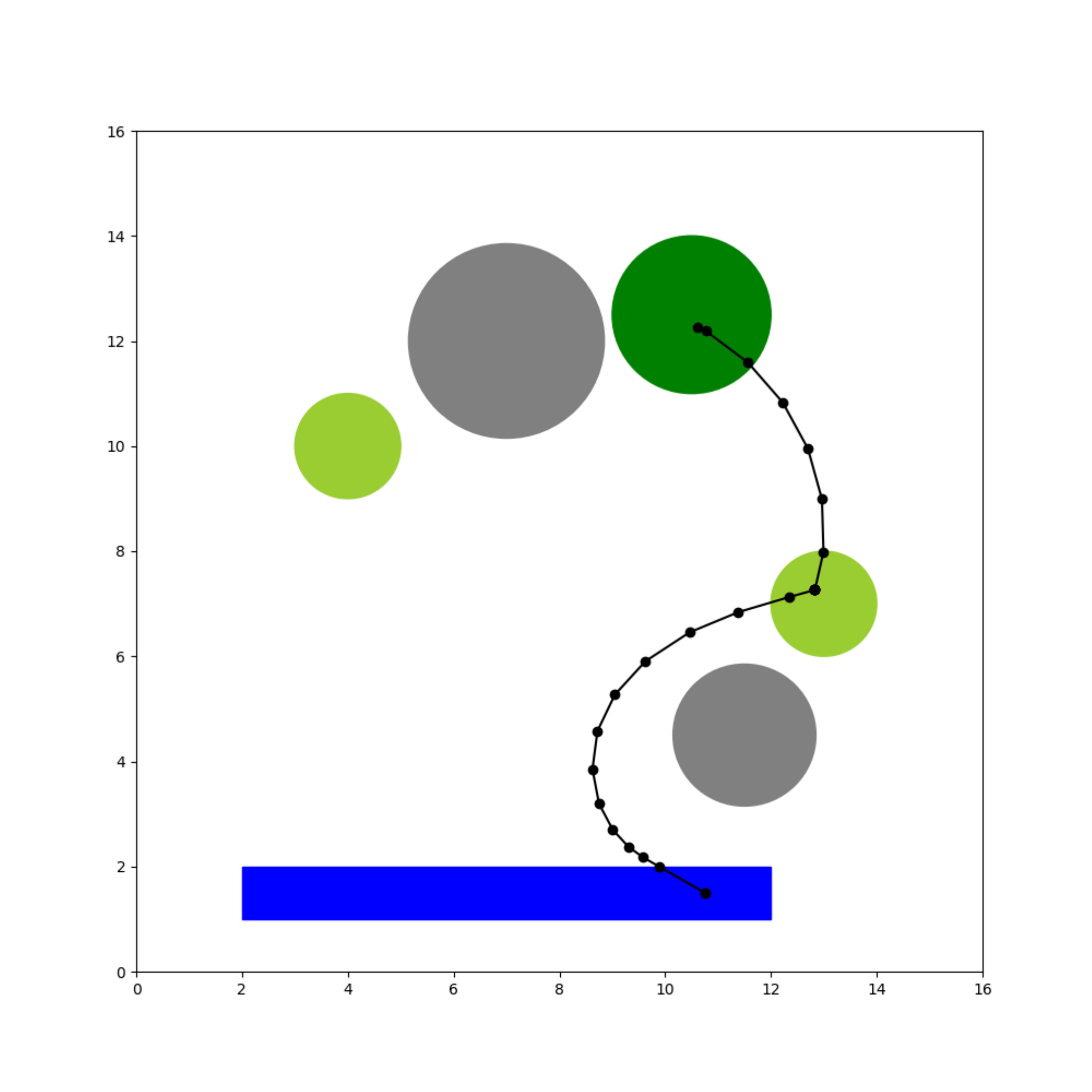}
    \captionof*{figure}{\small (a) Cluster 1}
    \label{fig:opt-a}
\end{minipage}
\begin{minipage}[b]{0.24\textwidth}
    \centering
    \includegraphics[width=\linewidth]{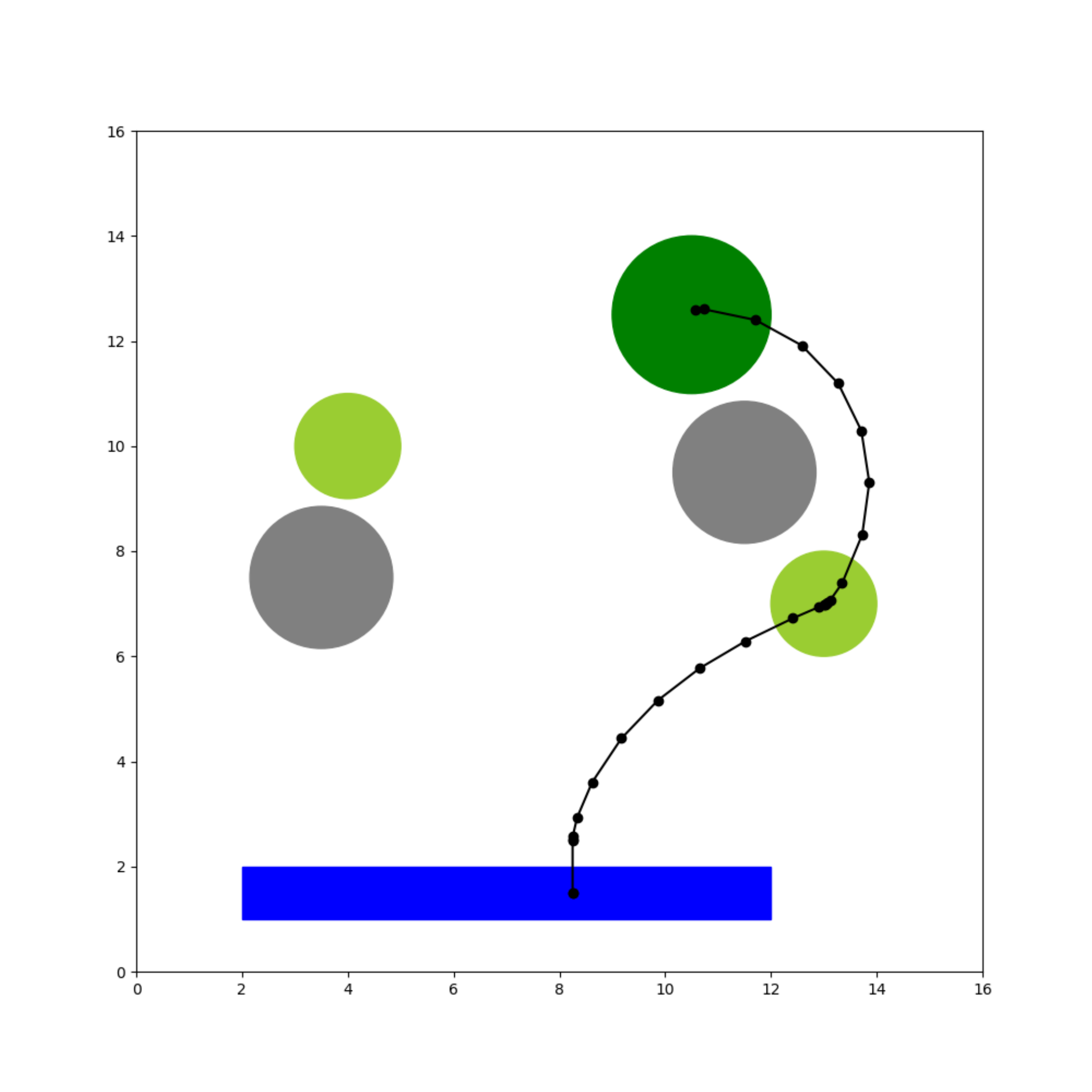}
    \captionof*{figure}{\small (b) Cluster 1}
    \label{fig:opt-b}
\end{minipage}
\begin{minipage}[b]{0.24\textwidth}
    \centering
    \includegraphics[width=\linewidth]{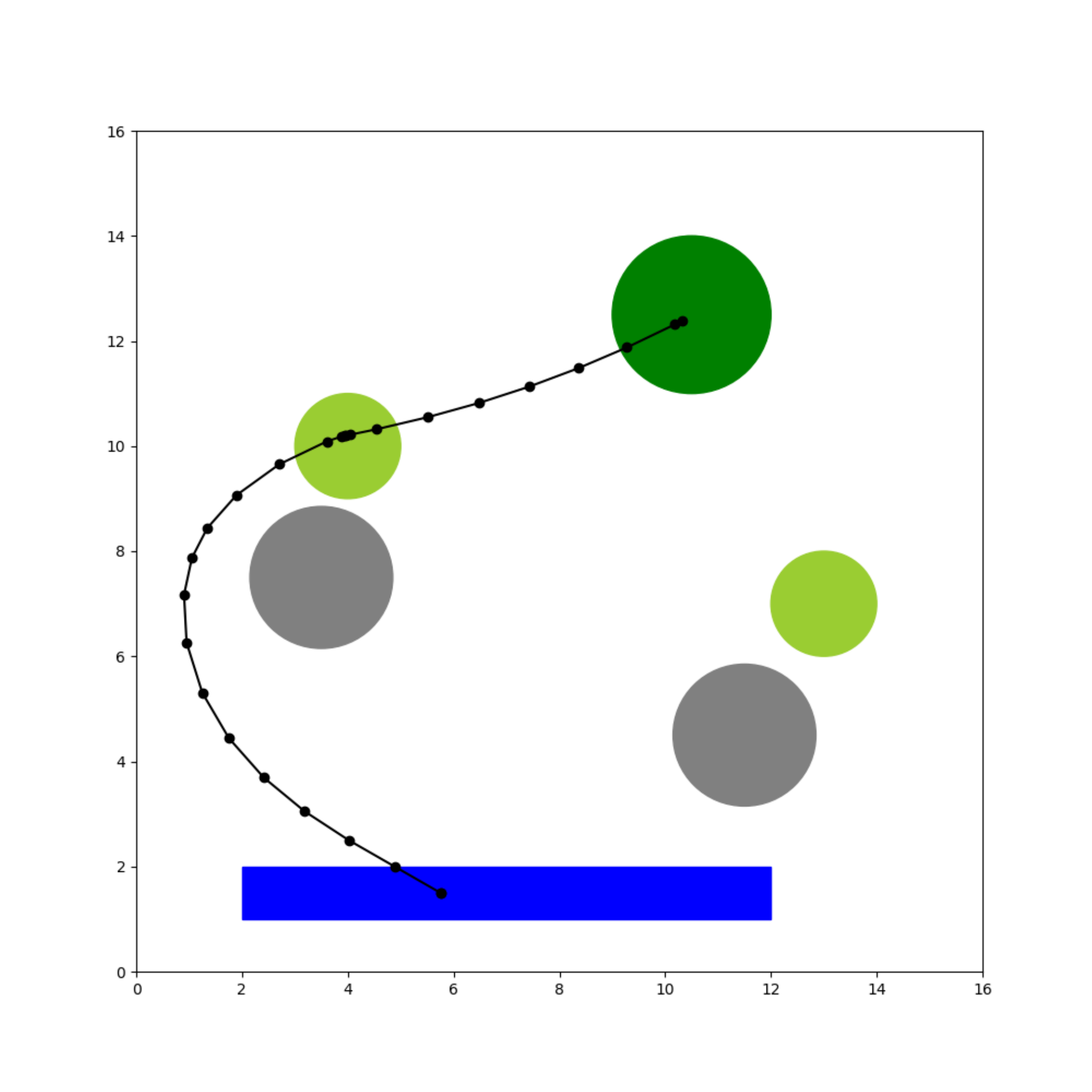}
    \captionof*{figure}{\small (c) Cluster 2}
    \label{fig:opt-c}
\end{minipage}
\begin{minipage}[b]{0.24\textwidth}
    \centering
    \includegraphics[width=\linewidth]{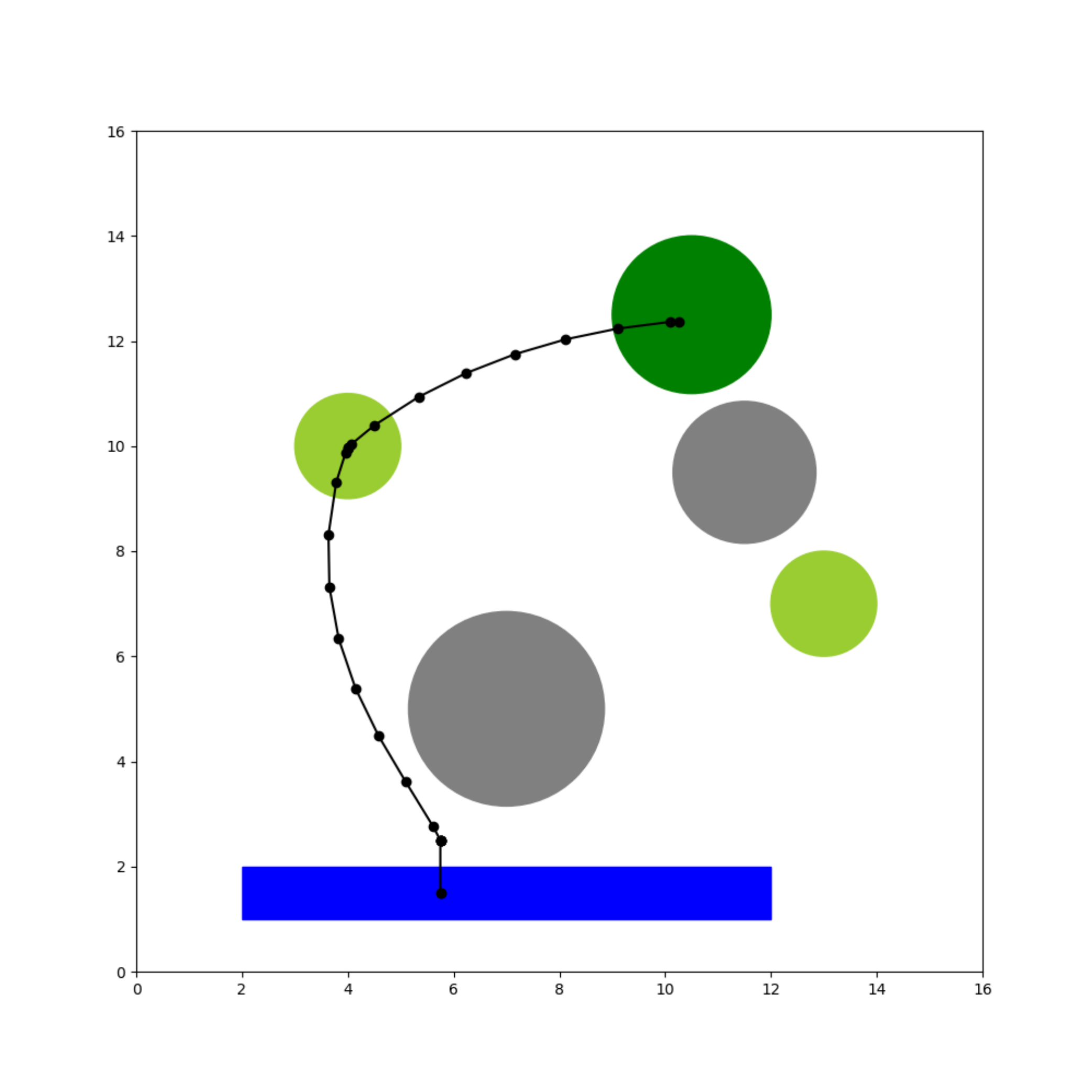}
    \captionof*{figure}{\small (d) Cluster 2}
    \label{fig:opt-d}
\end{minipage}

\vspace{1em}

% --- 2段目 ---
\begin{minipage}[b]{0.24\textwidth}
    \centering
    \includegraphics[width=\linewidth]{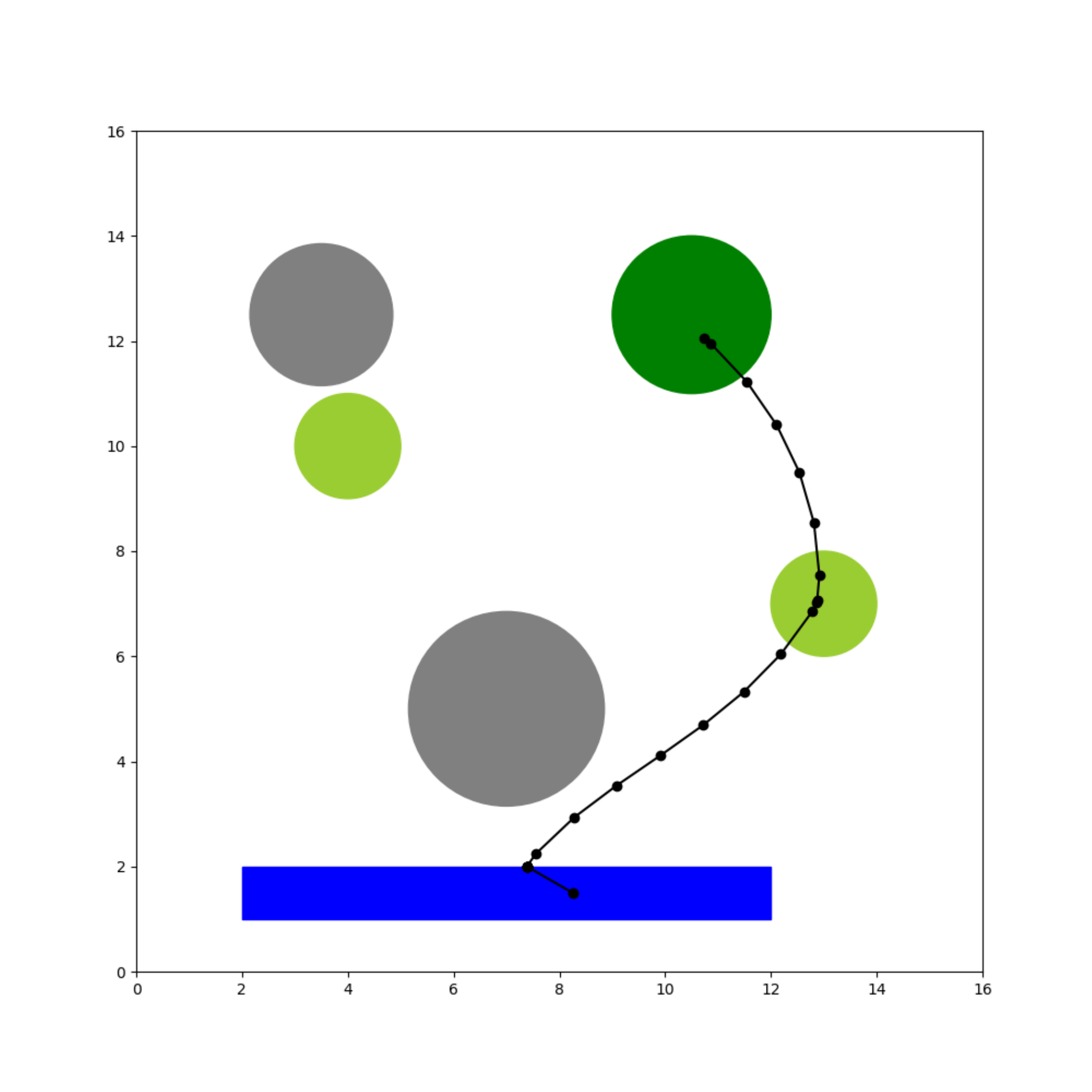}
    \captionof*{figure}{\small (e) Cluster 3}
    \label{fig:opt-e}
\end{minipage}
\begin{minipage}[b]{0.24\textwidth}
    \centering
    \includegraphics[width=\linewidth]{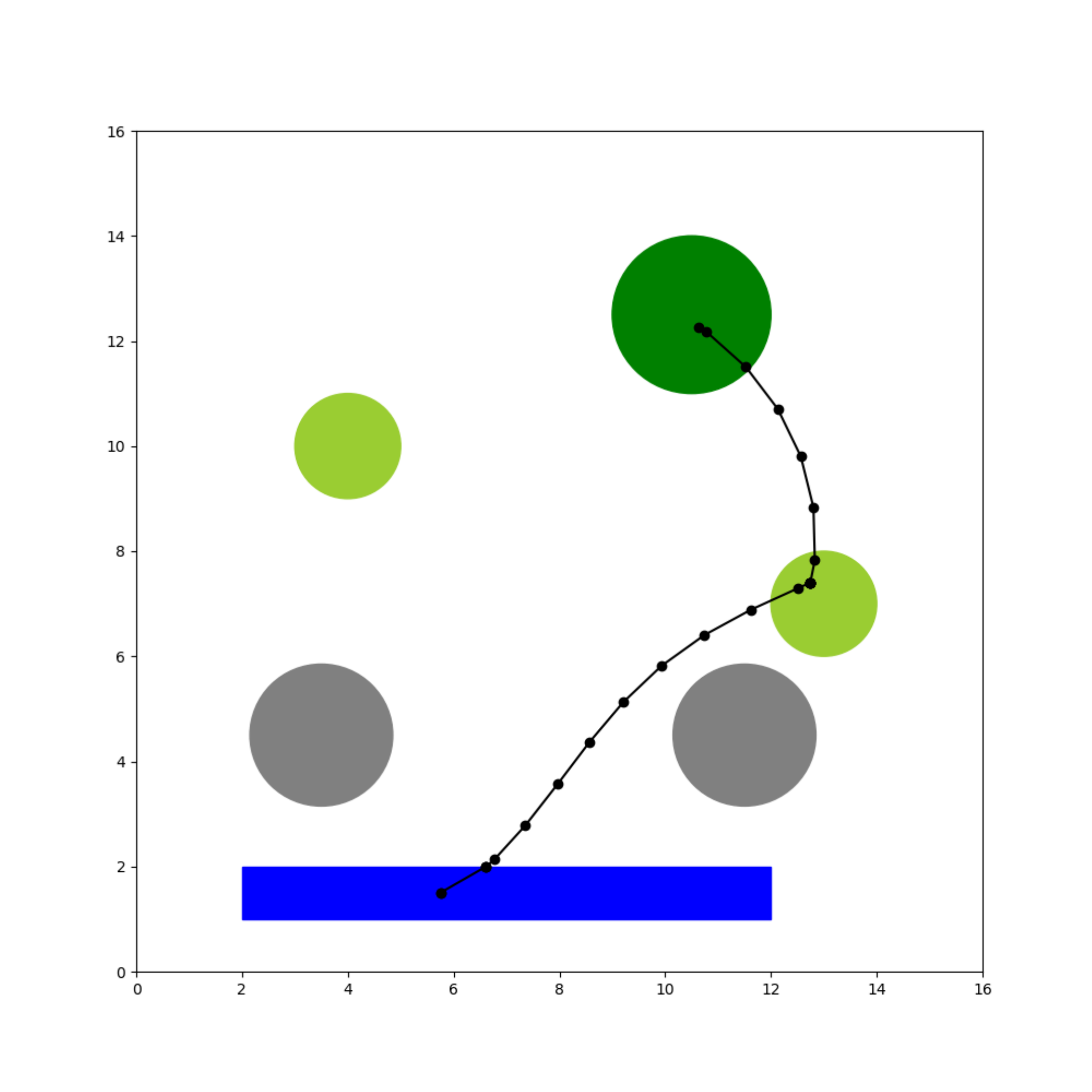}
    \captionof*{figure}{\small (f) Cluster 3}
    \label{fig:opt-f}
\end{minipage}
\begin{minipage}[b]{0.24\textwidth}
    \centering
    \includegraphics[width=\linewidth]{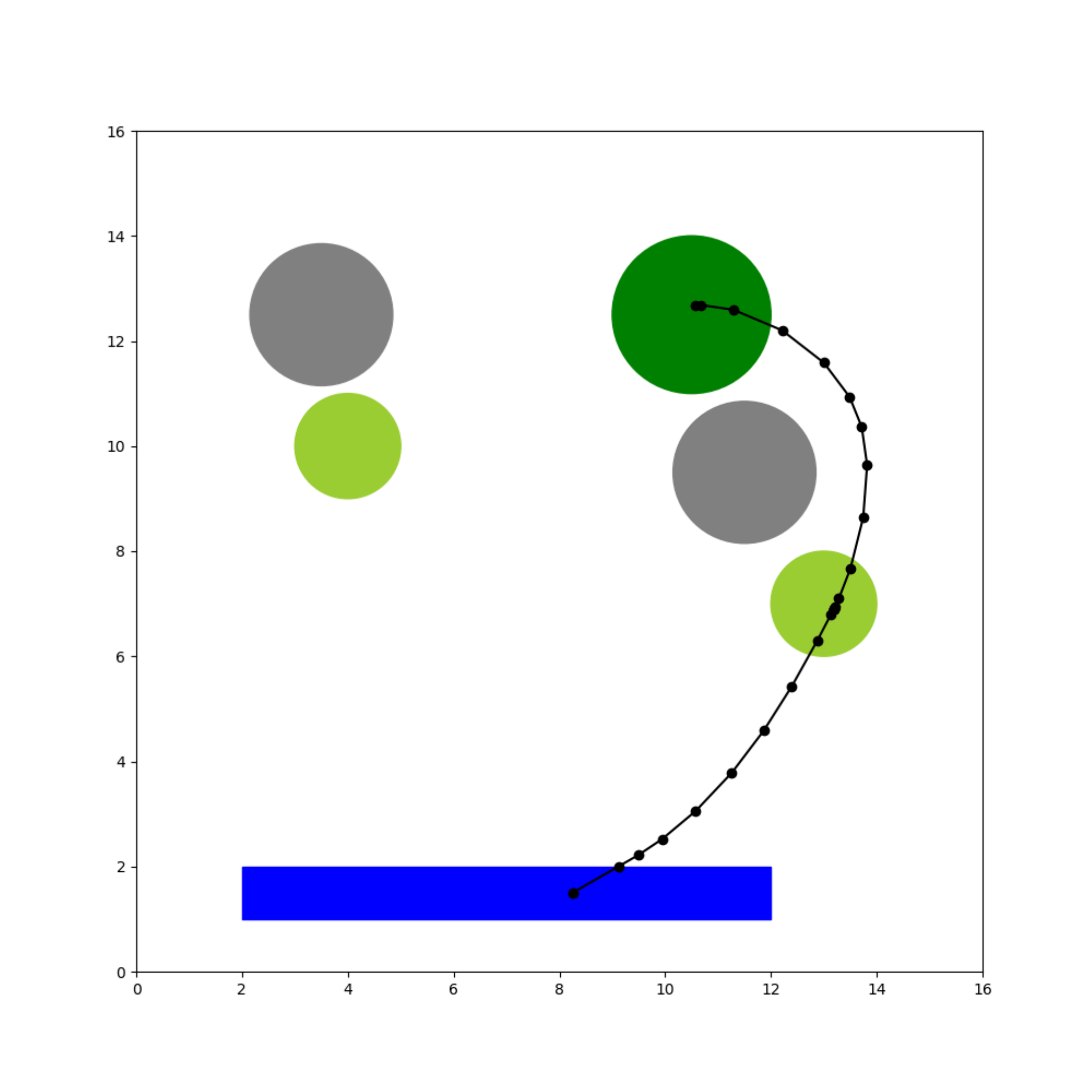}
    \captionof*{figure}{\small (g) Cluster 4}
    \label{fig:opt-g}
\end{minipage}
\begin{minipage}[b]{0.24\textwidth}
    \centering
    \includegraphics[width=\linewidth]{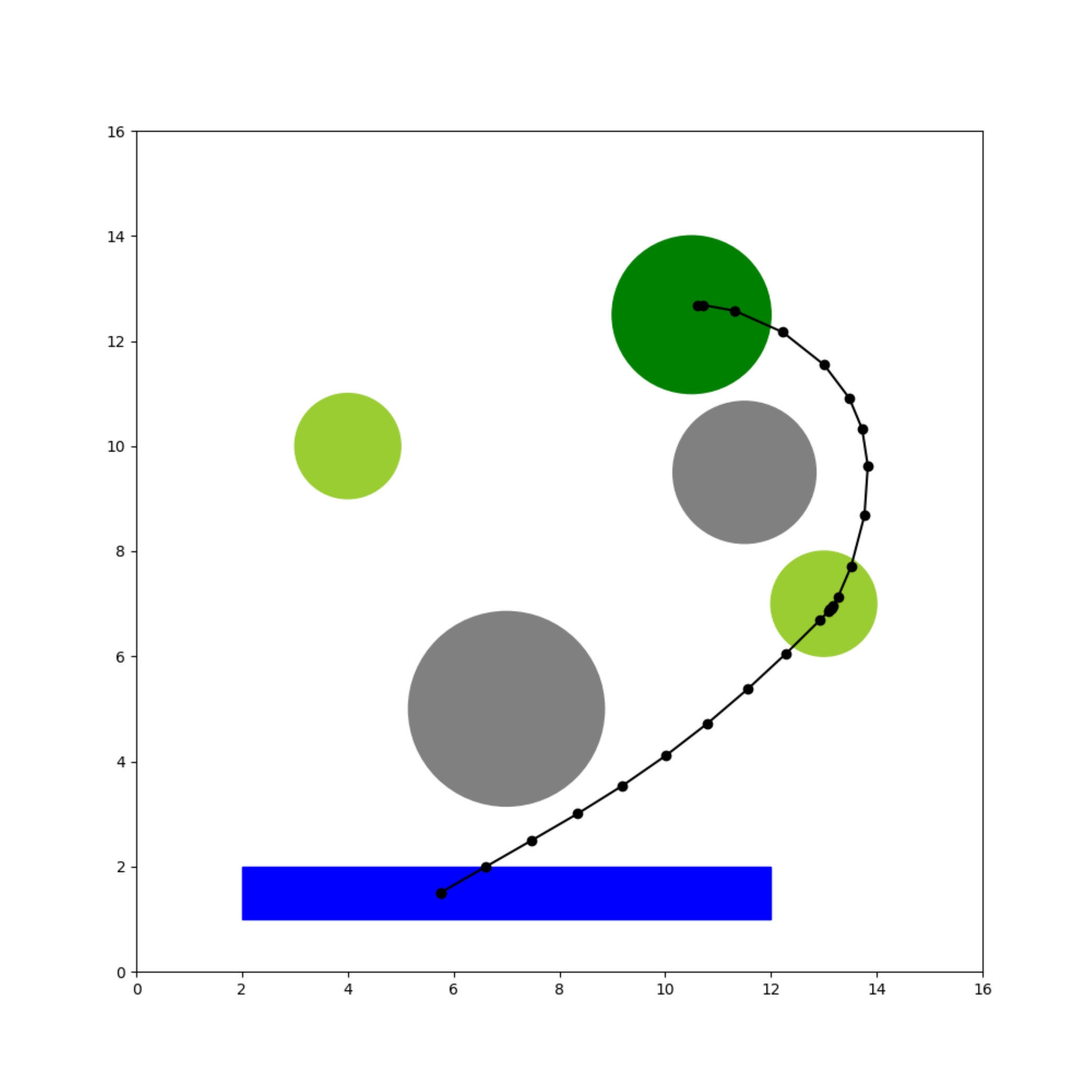}
    \captionof*{figure}{\small (h) Cluster 4}
    \label{fig:opt-h}
\end{minipage}

\caption{Optimal trajectories obtained by solving the optimization problem, classified into Cluster~1, Cluster~2, Cluster~3, and Cluster~4 for (a)--(b), (c)--(d), (e)--(f), and (g)--(h), respectively.}
\label{optimal_trajectory}
\end{figure*}

\subsection{Experiment Results and discussion}

\subsubsection{Clustering optimal trajectories}

Fig. \ref{optimal_trajectory} represents a part of the optimal trajectory obtained by solving the optimization problem. It should be noted that these trajectories are not generated by the neural network controller, but obtained by solving the optimization problem in \eqref{eq:optimization}. The number of clusters determined by the X-means method is 4. Fig. \ref{optimal_trajectory} (a)-(b), (c)-(d), (e)-(f), and (g)-(h) illustrate the trajectories categorized into Cluster 1, Cluster 2, Cluster 3, and Cluster 4, respectively. It can be observed that similar optimal trajectories tend to be grouped into the same class. %The classification network is trained by these clustering labels, with algorithm 2.

\subsubsection{Controllers Training and Performance Evaluation}

For comparison, controller training was conducted under two conditions: one employing clustering, as in the proposed method, and the other without clustering, as in the existing method. 

The training times are shown in Table~\ref{time}. In both the clustering-based and the single-controller settings, the number of training data was set to \(N_{\text{train}} = 48000\), and all controllers were trained for 30 epochs.  
For the proposed method, the four controllers were trained separately based on the clustering results, and the table includes the training time for each controller individually, along with the total training time. In addition, the training time for the single controller is also included for comparison. A comparison of the training times in Table~\ref{time} shows that the proposed method enables more efficient learning by utilizing similar trajectory data for training each controller. As a result, the total training time is reduced compared to that of training a single controller.

We prepared \(N_{\text{test}}=200\) test data, where both the initial states and obstacle parameters are randomly generated. As performance metrics for evaluating the controllers, we define the accuracy, the mean robustness loss, the mean control loss, the mean total loss and the mean cumulative distance-based cost to the goal.

The accuracy, denoted as \(\mathcal{A}\), quantifies the proportion of trajectories that satisfy the STL task specified in (\ref{STLtask}), and is defined as
\begin{align}
\mathcal{A} = \frac{1}{N_{\text{test}}} \sum_{j=1}^{N_{\text{test}}} 1\left[\rho^{\phi_{\Xi_j}}(x_{0:T}^{j}) > 0\right], \label{eq:accuracy}
\end{align}
where the indicator function \(1[\cdot]\) is defined as
\begin{align}
1\left[\rho^{\phi_{\Xi_j}}(x_{0:T}^{j}) > 0\right] =
\begin{cases}
1 & \text{if } \rho^{\phi_{\Xi_j}}(x_{0:T}^{j}) > 0, \\
0 & \text{otherwise}.
\end{cases} \label{eq:indicator}
\end{align}
We also define the mean robustness loss \(\overline{\mathcal{L}}_{\text{robustness}}\), the mean control loss \(\overline{\mathcal{L}}_{\text{control}}\), and the total loss \(\overline{\mathcal{L}}_{\text{total}}\)as follows: 
\begin{align}
\overline{\mathcal{L}}_{\text{total}} 
  &= -\overline{\mathcal{L}}_{\text{robustness}} + \gamma \overline{\mathcal{L}}_{\text{control}}, 
\end{align}  \begin{align}
\overline{\mathcal{L}}_{\text{robustness}} 
  &= \frac{1}{N_{\text{test}}} \sum_{j=1}^{N_{\text{test}}} \rho^{\phi_{\Xi_j}} \left( x_{0:T}^{j} \right), 
\end{align}  \begin{align}
  \overline{\mathcal{L}}_{\text{control}} 
  &= \frac{1}{|\mathcal{J}_{\text{succ}}|} \sum_{j_s \in \mathcal{J}_{\text{succ}}} \sum_{k=0}^{T-1} g(x_k^{j_s}, u_k^{j_s}),
\end{align}
where \(\mathcal{J}_{\text{succ}}\) denotes the index set of test data for which both the clustering-based (proposed) controller and the single controller successfully satisfy the STL specification, i.e., 
\begin{align*}
& \mathcal{J}_{\text{succ}} = \Big\{ j_s \in \{1, \dots, N_{\text{test}}\} \ \Big| 
\notag \\
& \quad \rho^{\phi_{\Xi_{j_s}}}(x^{j_s}_{0:T,\text{(clustering-based)}}) > 0 \  \wedge \
\rho^{\phi_{\Xi_{j_s}}}(x^{j_s}_{0:T,\text{(single)}}) > 0 \Big\}.
\end{align*}
Note that the control loss is evaluated only for the subset of test cases in which both controllers satisfy the STL task, thereby excluding failed trajectories whose control cost would be irrelevant in the context of STL satisfaction.
These losses follow a similar structure to the training losses defined in \eqref{loss_controlpolicy}--\eqref{eq:L_control},  
where the loss for each controller is denoted by \(\mathcal{L}_{\mathrm{total}, \ell}\).  
 The computed total loss \(\overline{\mathcal{L}}_{\text{total}}\) represents an approximation of the expectation in \eqref{eq:ocp}. 
The mean cumulative distance-based cost to the goal, denoted as \(\overline{\mathcal{C}}_{\text{dist}}\), is defined as:
\begin{align}
\overline{\mathcal{C}}_{\text{dist}} 
&= \frac{1}{N_{\text{test}}} \sum_{j=1}^{N_{\text{test}}} \sum_{k=0}^{T-1} \| x_{k+1}^j - x_{\text{goal}} \|_Q,
\label{C_goal}
\end{align}
where $x_{\text{goal}} = [p^{\text{goal}}_x,p^{\text{goal}}_y,0,0,0]^\top$ and
\begin{align}
Q &= 
\begin{bmatrix}
1 & 0 & 0 & 0 & 0 \\
0 & 1 & 0 & 0 & 0 \\
0 & 0 & 0 & 0 & 0 \\
0 & 0 & 0 & 0 & 0 \\
0 & 0 & 0 & 0 & 0 
\end{bmatrix}.
\end{align}
Thus, a smaller \(\mathcal{C}_{\text{dist}}\) indicates that the trajectory reaches the goal more efficiently, whereas a larger value suggests a longer or less efficient trajectory.

\begin{comment}
    
\begin{figure*}[t]
\centering
    \subfigure[Case 1]{%
        \includegraphics[clip, width=0.5\columnwidth]{Figs/Results/a-96.pdf}}%
    \subfigure[Case 2]{%
        \includegraphics[clip, width=0.5\columnwidth]{Figs/Results/b-3.pdf}}%
  \subfigure[Case 3]{%
        \includegraphics[clip, width=0.5\columnwidth]{Figs/Results/c-176.pdf}}
  \subfigure[Case 4]
  {%
        \includegraphics[clip, width=0.5\columnwidth]{Figs/Results/d-137.pdf}}%
        \par
    \subfigure[Case 5]{%
        \includegraphics[clip, width=0.5\columnwidth]{Figs/Results/e-19.pdf}}%
  \subfigure[Case 6]{%
        \includegraphics[clip, width=0.5\columnwidth]{Figs/Results/f-181.pdf}}
  \subfigure[Case 7]{%
        \includegraphics[clip, width=0.5\columnwidth]{Figs/Results/g-11.pdf}}%
  \subfigure[Case 8]{%
		\includegraphics[clip, width=0.5\columnwidth]{Figs/Results/h-74.pdf}}
    \caption{Trajectories by applying the clustering-based (proposed) approach (red) and the single RNN approach (black). }
\label{final_result}
\end{figure*}
\end{comment}

\begin{figure*}[t]
\centering

% 1段目
\begin{minipage}[b]{0.24\textwidth}
    \centering
    \includegraphics[width=\linewidth]{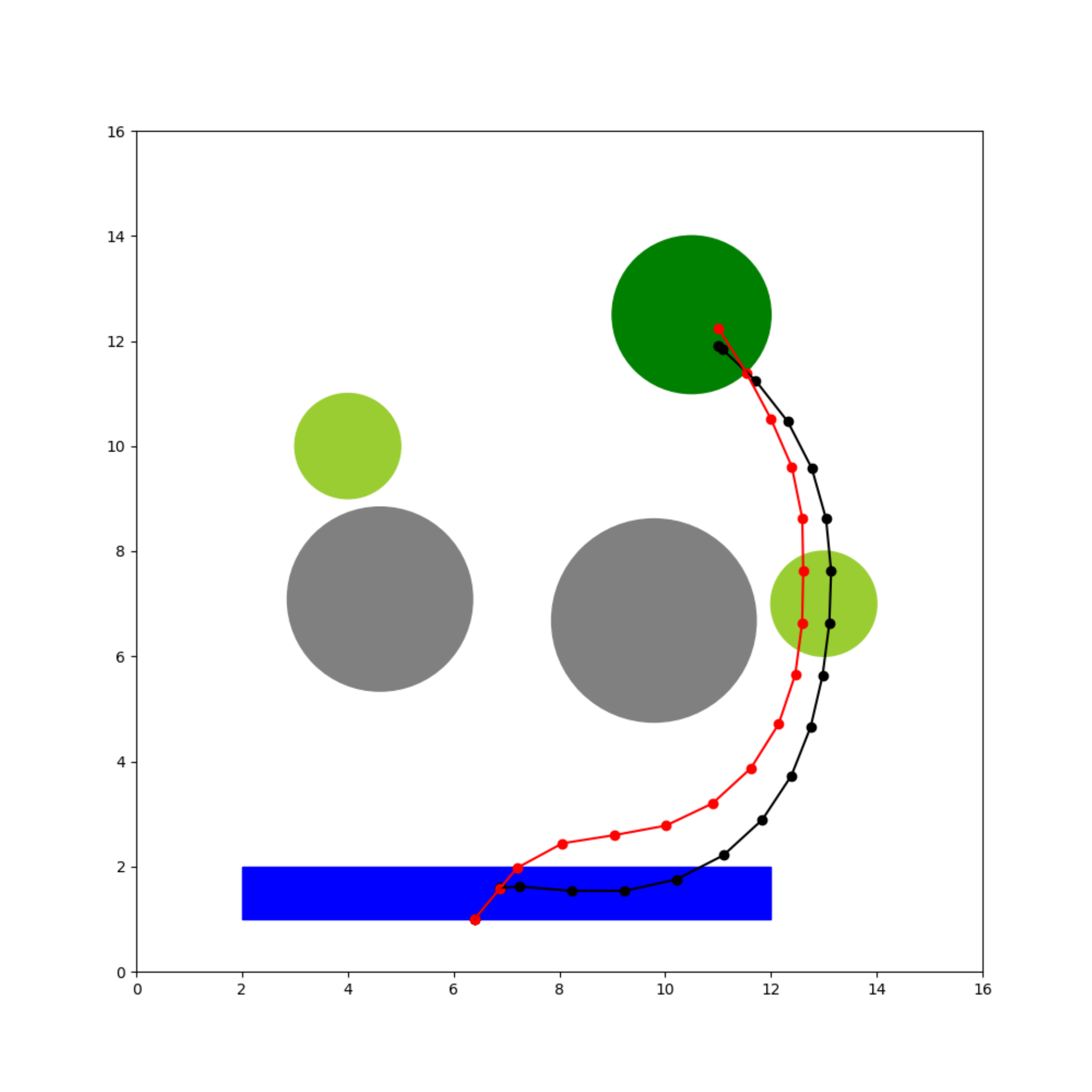}
    \captionof*{figure}{(a) Case 1}
    \label{fig:a}
\end{minipage}
\begin{minipage}[b]{0.24\textwidth}
    \centering
    \includegraphics[width=\linewidth]{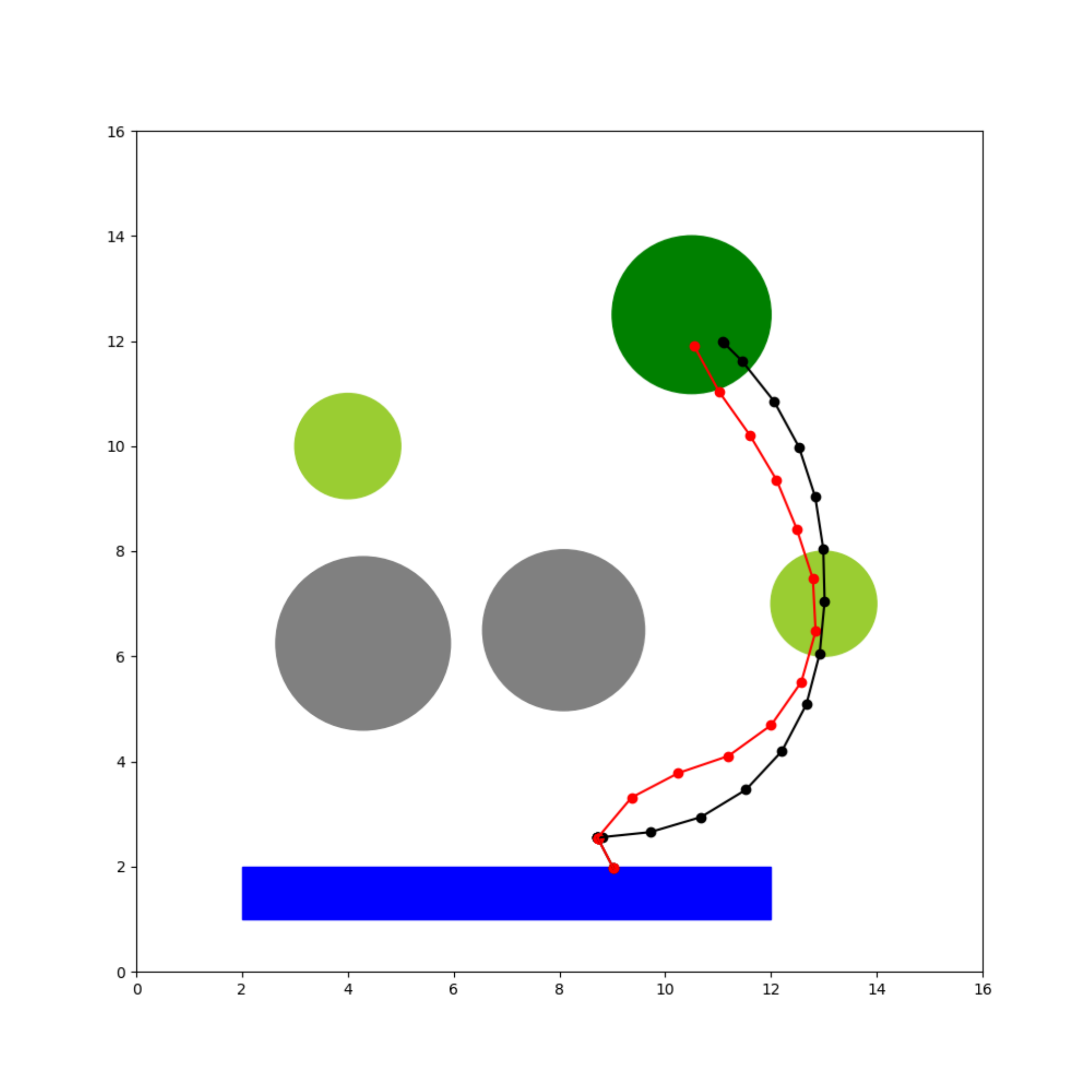}
    \captionof*{figure}{(b) Case 2}
    \label{fig:b}
\end{minipage}
\begin{minipage}[b]{0.24\textwidth}
    \centering
    \includegraphics[width=\linewidth]{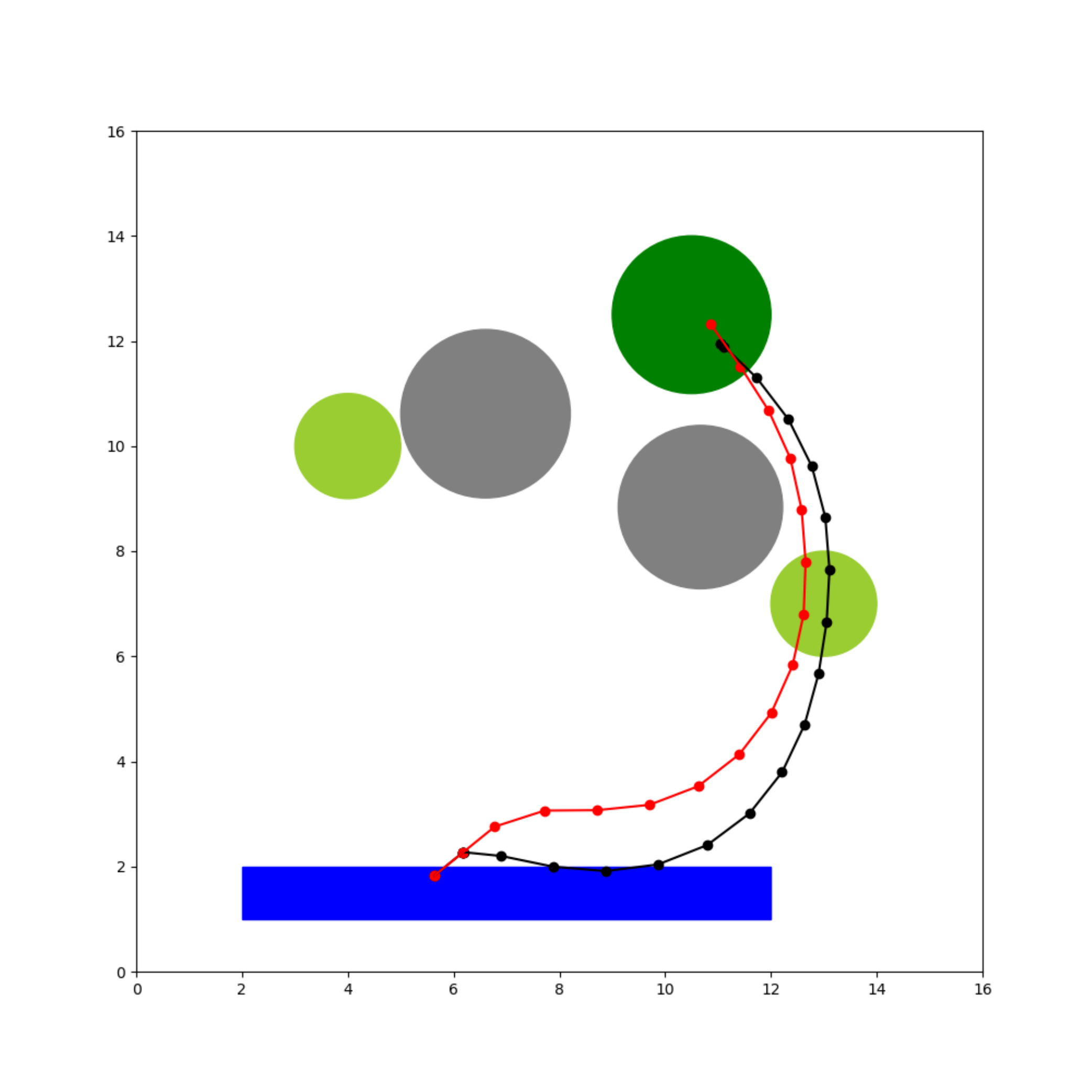}
    \captionof*{figure}{(c) Case 3}
    \label{fig:c}
\end{minipage}
\begin{minipage}[b]{0.24\textwidth}
    \centering
    \includegraphics[width=\linewidth]{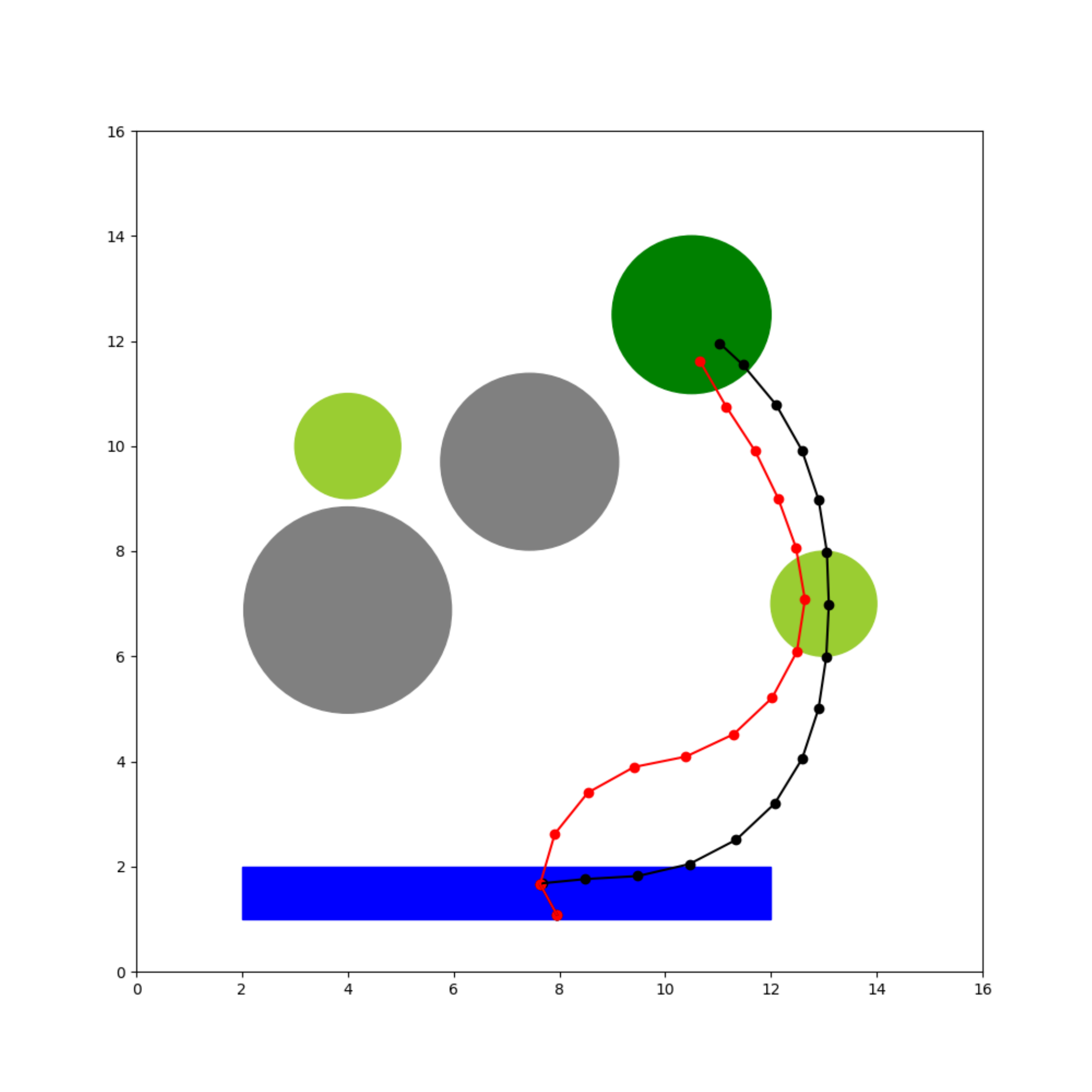}
    \captionof*{figure}{(d) Case 4}
    \label{fig:d}
\end{minipage}

\vspace{1em}

% 2段目
\begin{minipage}[b]{0.24\textwidth}
    \centering
    \includegraphics[width=\linewidth]{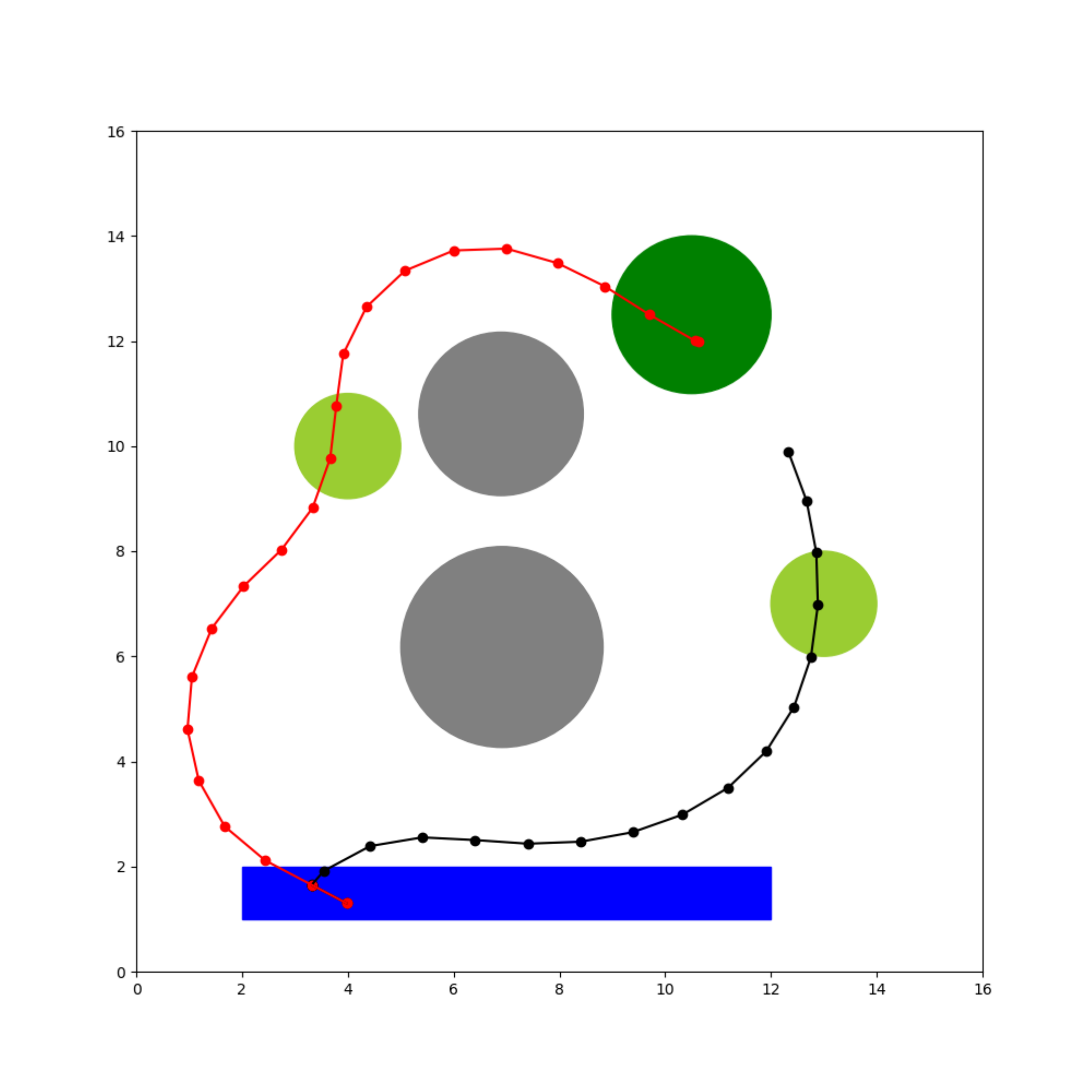}
    \captionof*{figure}{(e) Case 5}
    \label{fig:e}
\end{minipage}
\begin{minipage}[b]{0.24\textwidth}
    \centering
    \includegraphics[width=\linewidth]{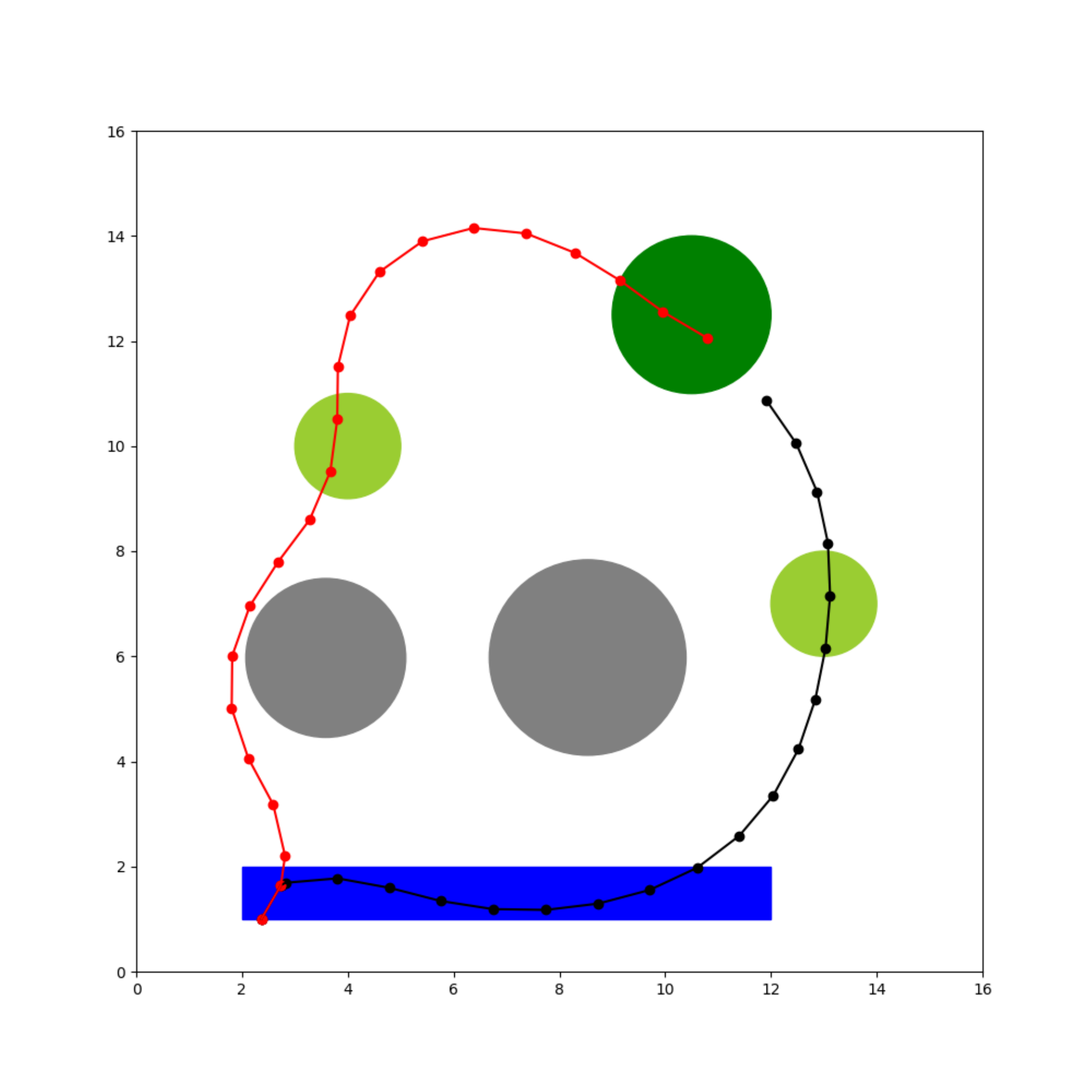}
    \captionof*{figure}{(f) Case 6}
    \label{fig:f}
\end{minipage}
\begin{minipage}[b]{0.24\textwidth}
    \centering
    \includegraphics[width=\linewidth]{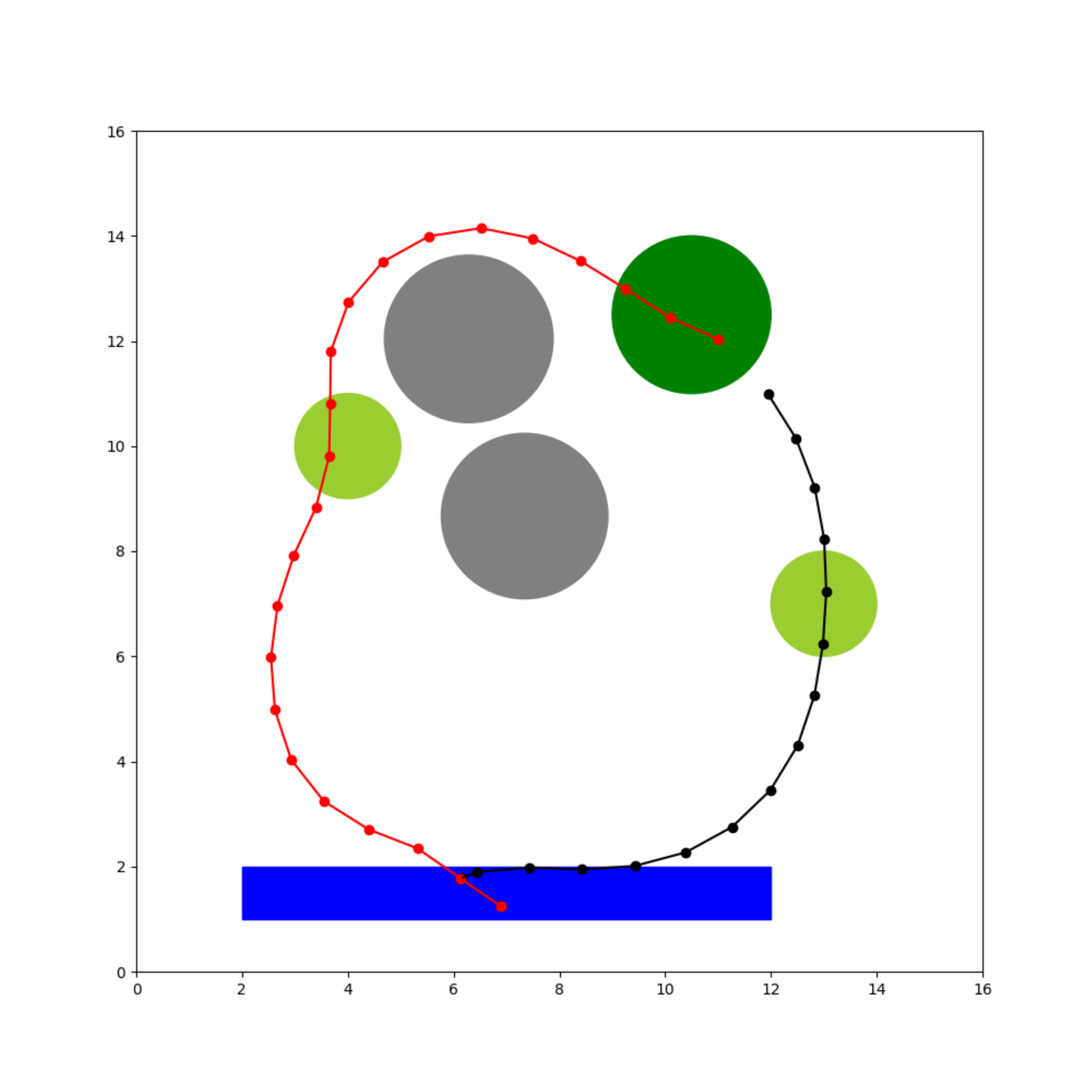}
    \captionof*{figure}{(g) Case 7}
    \label{fig:g}
\end{minipage}
\begin{minipage}[b]{0.24\textwidth}
    \centering
    \includegraphics[width=\linewidth]{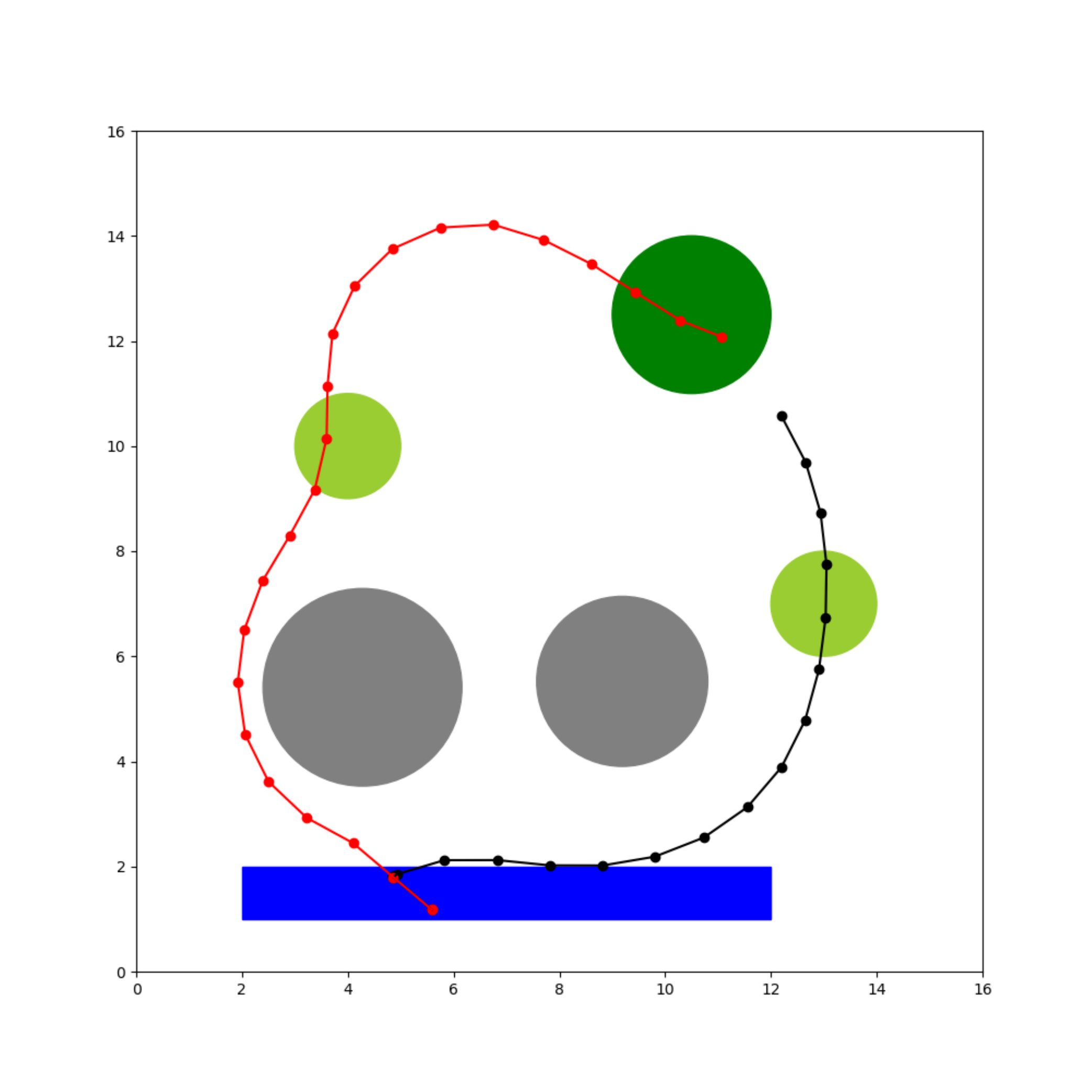}
    \captionof*{figure}{(h) Case 8}
    \label{fig:h}
\end{minipage}

\caption{Trajectories by applying the clustering-based (proposed) approach (red) and the single RNN approach (black).}
\label{final_result}
\end{figure*}

%以降の記載を結果に合わせて変更予定

Table~\ref{score} shows the overall performance metrics obtained from the test dataset, comparing the single RNN-based controller and the proposed clustering-based RNN controllers. The results show that the proposed method achieves higher accuracy and lower mean losses, including the mean cumulative distance-based cost to the goal. Fig.~\ref{final_result} shows several test cases. The red trajectories correspond to those generated by the proposed clustering-based controllers, while the black trajectories correspond to those generated by the single controller. The detailed metrics for each case are listed in Table~\ref{tab:casewise_scores}.

Based on these results, we can make the following observations:
\begin{itemize}
    \item As seen in the black trajectories in Fig.~\ref{final_result}, the single controller causes all test trajectories to pass through the same transit region toward the goal, regardless of the initial states. Consequently, some trajectories follow a longer and inefficient path, resulting in failure to reach the goal region within the given horizon.
    \item In contrast, the red trajectories in Fig.~\ref{final_result} take different transit regions and routes depending on the initial states and obstacle parameters. As a result, the trajectories reach the goal region within the given horizon.
    \item As shown in Table~\ref{score}, the proposed method improves all evaluation metrics. These results suggest that the proposed method enables the generation of more accurate and efficient trajectories.
\end{itemize}

\begin{table}[t]
  \centering
  \caption{Comparison of test performance between the single RNN controller and the proposed clustering-based RNN controllers.}
  \renewcommand{\arraystretch}{1.5} % 行間の調整
  \begin{tabular}{ccc} 
    \hline
    Metric & Single RNN & Clustering-based RNN \\ 
    \hline \hline
    Accuracy \(\mathcal{A}\) & 0.800 & 0.915\\
    Mean robustness loss \(\overline{\mathcal{L}}_{\text{robustness}}\) &0.304  & 0.457\\
    Mean control loss \(\overline{\mathcal{L}}_{\text{control}}\) & 54.09 & 49.55\\
    Mean total loss \(\overline{\mathcal{L}}_{\text{total}}\) & $15.92$ & $14.41$\\
    Mean distance cost \(\overline{\mathcal{C}}_{\text{dist}} 
\) & 175.4 & 137.0\\
    \hline
  \end{tabular}
  \label{score}
\end{table}

\begin{table}[t]
  \centering
  \caption{Test scores for each case and controller type}
  \renewcommand{\arraystretch}{1.3}
  \begin{tabular}{cccccc}
    \hline
    Case & Approach & $\mathcal{L}_{\text{robustness}}$ & $\mathcal{L}_{\text{control}}$ & $\mathcal{L}_{\text{total}}$ & $\mathcal{C}_{\text{dist}}$ \\
    \hline \hline
    \multirow{2}{*}{Case 1} & single            & 0.608 & 55.03 & 15.90 & 173.4 \\
                            & clustering-based  & 0.453 & 52.26 & 15.23 & 121.3 \\
    \hline
    \multirow{2}{*}{Case 2} & single            & 0.713 & 48.85 & 13.94 & 166.0 \\
                            & clustering-based  & 0.484 & 48.01 & 13.92 & 116.9 \\
    \hline
    \multirow{2}{*}{Case 3} & single            & 0.647 & 55.80 & 16.09 & 169.2 \\
                            & clustering-based  &0.360 & 52.88 & 15.50 & 116.6 \\
    \hline
    \multirow{2}{*}{Case 4} & single            & 0.736 & 54.77 & 15.70 & 187.8 \\
                            & clustering-based  & 0.597 & 50.32 & 14.50 & 123.4 \\
    \hline
    \multirow{2}{*}{Case 5} & single            & $-1.684$ & - & - & 225.0 \\
                            & clustering-based  & 0.594 & - & - & 173.7 \\
    \hline
    \multirow{2}{*}{Case 6} & single            & $-0.658$ & - & - & 221.4 \\
                            & clustering-based  & 0.232 & - & - & 178.8 \\
    \hline
    \multirow{2}{*}{Case 7} & single            & $-0.589$ & - & - & 206.4 \\
                            & clustering-based  & 0.501 & - & - & 167.4 \\
    \hline
    \multirow{2}{*}{Case 8} & single            & $-1.066$ & - & - & 214.2 \\
                            & clustering-based  & 0.481 & - & - & 168.6 \\
    \hline
  \end{tabular}
  \label{tab:casewise_scores}
\end{table}

\begin{comment}
  \begin{table*}[h]
  \centering
  \caption{Test scores for each case and controller type}
  \renewcommand{\arraystretch}{1.3}
  \begin{tabular}{cccccc || cccccc}
    \hline
    Case & Method & $\mathcal{L}_{\text{robustness}}$ & $\mathcal{L}_{\text{control}}$ & $\mathcal{L}_{\text{total}}$ & $\mathcal{C}_{\text{dist}}$ &
    Case & Method & $\mathcal{L}_{\text{robustness}}$ & $\mathcal{L}_{\text{control}}$ & $\mathcal{L}_{\text{total}}$ & $\mathcal{C}_{\text{dist}}$
    \\
    \hline \hline
    \multirow{2}{*}{Case 1} & single            & - & - & - & - &                       
    \multirow{2}{*}{Case 2} & single            & - & - & - & - \\
    & clustering-based  & - & - & - & - &
    & clustering-based  & - & - & - & - \\
    \hline
    \multirow{2}{*}{Case 3} & single            & - & - & - & - &                       
    \multirow{2}{*}{Case 4} & single            & - & - & - & - \\
    & clustering-based  & - & - & - & - &
    & clustering-based  & - & - & - & - \\
    \hline
    \multirow{2}{*}{Case 5} & single            & - & - & - & - &                       
    \multirow{2}{*}{Case 6} & single            & - & - & - & - \\
    & clustering-based  & - & - & - & - &
    & clustering-based  & - & - & - & - \\
    \hline
    \multirow{2}{*}{Case 7} & single            & - & - & - & - &                       
    \multirow{2}{*}{Case 8} & single            & - & - & - & - \\
    & clustering-based  & - & - & - & - &
    & clustering-based  & - & - & - & - \\
    \hline
  \end{tabular}
  \label{tab:casewise_scores}
\end{table*}

\end{comment}

\section{Conclusion}
This paper presented a framework to improve the control performance of recurrent neural network (RNN)-based controllers under Signal Temporal Logic (STL) specifications. Unlike existing methods that rely on a single neural network, the proposed approach accounts for environmental variations, specifically changes in initial states and obstacle parameters assumed in this study, by clustering optimal trajectories and assigning a specialized controller to each cluster. To enable adaptive controller selection, a classification network is trained to map new environmental conditions (i.e., initial states and obstacle parameters) to the corresponding cluster.

Based on the similarity of optimal trajectories, the proposed method clusters initial states and obstacle parameters, and assigns a specialized controller to each cluster using a classification network. Each controller is then trained individually.

In numerical experiments on a dynamic vehicle path planning problem, the proposed method improved all evaluation metrics, including the STL satisfaction rate, robustness loss, control loss, and cumulative distance-based cost, compared to the existing approach. These results demonstrate the effectiveness of selecting appropriate controllers based on environmental conditions and initial states to satisfy STL specifications.

\appendices
\section {Criterion for Cluster Number Selection in X-means}
\label{appendix:clustering}
This appendix provides the formulation and analysis of the evaluation criterion used in the X-means algorithm for dynamically determining the number of clusters.  
While the main text briefly introduced AIC and BIC as model selection metrics, here we present their explicit mathematical definitions and describe a new formulation that integrates them.  
This study employs a single evaluation criterion, the Mixed Information Criterion (MIC), which combines AIC and BIC in a weighted formulation.

To implement the clustering process, we used the PyClustering library~\cite{pyclustering}, which supports both AIC and BIC as evaluation criteria.  
AIC and BIC are defined as follows:
\begin{align}
\mathrm{AIC}(\mathcal{M}) &= -\mathcal{L}(\mathcal{D}) + d, \\
\mathrm{BIC}(\mathcal{M}) &= -\mathcal{L}(\mathcal{D}) + \frac{d}{2} \log (\mathcal{R}),
\end{align}
where \(\mathcal{L}(\mathcal{D})\) denotes the log-likelihood of dataset \(\mathcal{D}\) under model \(\mathcal{M}\), with \(d\) parameters and \(\mathcal{R}\) samples.

AIC evaluates models based on their likelihood, applying a penalty proportional to the number of parameters. Because of its relatively smaller penalty, AIC tends to favor more complex models.  
In contrast, BIC imposes a stronger penalty that increases with the sample size, typically leading to simpler models as the dataset grows.

While both criteria are useful, they exhibit different selection behaviors. To balance these characteristics, we introduce the Mixed Information Criterion (MIC), defined as:
\begin{align}
\Psi(\mathcal{M}) = -\mathcal{L}(\mathcal{D}) + \gamma_c \cdot \frac{d}{2} \log (\mathcal{R}) + (1 - \gamma_c) d,
\end{align}
where \(\gamma_c \in [0,1]\) is a weighting coefficient that interpolates between BIC (\(\gamma_c = 1\)) and AIC (\(\gamma_c = 0\)).

The parameter \(\gamma_c\) can be either treated as a hyperparameter or determined dynamically based on the structure of the data.  
To reflect cluster separability, we define \(\gamma_c\) using the within-cluster and between-cluster variances:
\begin{align}
\gamma_c &= \frac{\mathcal{V}_b}{\mathcal{V}_b + \mathcal{V}_w}, \\
\mathcal{V}_w &= \sum_{p=1}^{\mathcal{K}} \sum_{\chi \in C_p} \| \chi - \mu_{\text{cluster},p} \|^2, \\
\mathcal{V}_b &= \sum_{p=1}^{\mathcal{K}} N_p \| \mu_{\text{cluster},p} - \mu_{\text{all}} \|^2,
\end{align}
where \(\mathcal{V}_w\) and \(\mathcal{V}_b\) are the within-cluster and between-cluster variances, respectively.  
Here, \(C_p\) is the set of data points in cluster \(p\), \(N_p\) is its size, \(\mu_{\text{cluster},p}\) is its centroid, and \(\mu_{\text{all}}\) is the overall mean of the dataset.

Variance-based formulations are commonly used in clustering evaluation metrics, including but not limited to the established Calinski–Harabasz index~\cite{Calinski–Harabasz}, and have been shown to be effective for assessing cluster separability.
When clusters are well-separated, \(\mathcal{V}_b\) dominates and \(\gamma_c\) approaches 1, making MIC behave more like BIC. In such cases, selecting a simpler model helps reduce the risk of overfitting. Conversely, when clusters are not clearly separated, \(\mathcal{V}_w\) dominates and \(\gamma_c\) approaches 0, making MIC behave more like AIC, which favors more complex models that better capture overlapping structures.

This adaptive formulation enables the number of clusters to be determined in a data-driven manner, reflecting the structural complexity of the dataset.

\bibliographystyle{IEEEtran}
\bibliography{myrefs}

\begin{thebibliography}{10}
\providecommand{\url}[1]{#1}
\csname url@samestyle\endcsname
\providecommand{\newblock}{\relax}
\providecommand{\bibinfo}[2]{#2}
\providecommand{\BIBentrySTDinterwordspacing}{\spaceskip=0pt\relax}
\providecommand{\BIBentryALTinterwordstretchfactor}{4}
\providecommand{\BIBentryALTinterwordspacing}{\spaceskip=\fontdimen2\font plus
\BIBentryALTinterwordstretchfactor\fontdimen3\font minus \fontdimen4\font\relax}
\providecommand{\BIBforeignlanguage}[2]{{%
\expandafter\ifx\csname l@#1\endcsname\relax
\typeout{** WARNING: IEEEtran.bst: No hyphenation pattern has been}%
\typeout{** loaded for the language `#1'. Using the pattern for}%
\typeout{** the default language instead.}%
\else
\language=\csname l@#1\endcsname
\fi
#2}}
\providecommand{\BIBdecl}{\relax}
\BIBdecl

\bibitem{maler2004monitoring}
O.~Maler and D.~Nickovic, ``Monitoring temporal properties of continuous signals,'' in \emph{International symposium on formal techniques in real-time and fault-tolerant systems}.\hskip 1em plus 0.5em minus 0.4em\relax Springer, 2004, pp. 152--166.

\bibitem{donze2010robust}
A.~Donz{\'e} and O.~Maler, ``Robust satisfaction of temporal logic over real-valued signals,'' in \emph{International Conference on Formal Modeling and Analysis of Timed Systems}.\hskip 1em plus 0.5em minus 0.4em\relax Springer, 2010, pp. 92--106.

\bibitem{raman2014}
V.~Raman, A.~Donzé, M.~Maasoumy, R.~M. Murray, A.~Sangiovanni-Vincentelli, and S.~A. Seshia, ``Model predictive control with signal temporal logic specifications,'' in \emph{53rd IEEE Conference on Decision and Control}, 2014, pp. 81--87.

\bibitem{Sadraddini2015}
S.~Sadraddini and C.~Belta, ``Robust temporal logic model predictive control,'' in \emph{2015 53rd Annual Allerton Conference on Communication, Control, and Computing (Allerton)}, 2015, pp. 772--779.

\bibitem{pant2017smooth}
Y.~V. Pant, H.~Abbas, and R.~Mangharam, ``Smooth operator: Control using the smooth robustness of temporal logic,'' in \emph{2017 IEEE Conference on Control Technology and Applications (CCTA)}, 2017, pp. 1235--1240.

\bibitem{takayama2023}
Y.~Takayama, K.~Hashimoto, and T.~Ohtsuka, ``Signal temporal logic meets convex-concave programming: A structure-exploiting sqp algorithm for stl specifications,'' in \emph{2023 62nd IEEE Conference on Decision and Control (CDC)}, 2023, pp. 6855--6862.

\bibitem{leung2023backpropagation}
K.~Leung, N.~Ar{\'e}chiga, and M.~Pavone, ``Backpropagation through signal temporal logic specifications: Infusing logical structure into gradient-based methods,'' \emph{The International Journal of Robotics Research}, vol.~42, no.~6, pp. 356--370, 2023.

\bibitem{lindemann2019control}
L.~Lindemann and D.~V. Dimarogonas, ``Control barrier functions for multi-agent systems under conflicting local signal temporal logic tasks,'' \emph{IEEE control systems letters}, vol.~3, no.~3, pp. 757--762, 2019.

\bibitem{yu2024continuous}
P.~Yu, X.~Tan, and D.~V. Dimarogonas, ``Continuous-time control synthesis under nested signal temporal logic specifications,'' \emph{IEEE Transactions on Robotics}, vol.~40, pp. 2272--2286, 2024.

\bibitem{verginis2024planning}
C.~K. Verginis, Y.~Kantaros, and D.~V. Dimarogonas, ``Planning and control of multi-robot-object systems under temporal logic tasks and uncertain dynamics,'' \emph{Robotics and Autonomous Systems}, vol. 174, p. 104646, 2024.

\bibitem{liu2021recurrent}
W.~Liu, N.~Mehdipour, and C.~Belta, ``Recurrent neural network controllers for signal temporal logic specifications subject to safety constraints,'' \emph{IEEE Control Systems Letters}, vol.~6, pp. 91--96, 2021.

\bibitem{yaghoubi2019worst}
S.~Yaghoubi and G.~Fainekos, ``Worst-case satisfaction of stl specifications using feedforward neural network controllers: a lagrange multipliers approach,'' \emph{ACM Transactions on Embedded Computing Systems (TECS)}, vol.~18, no.~5s, pp. 1--20, 2019.

\bibitem{stlmpc}
Y.~Meng and C.~Fan, ``Signal temporal logic neural predictive control,'' \emph{IEEE Robotics and Automation Letters}, vol.~8, no.~11, pp. 7719--7726, 2023.

\bibitem{PuranicSTL}
A.~G. Puranic, J.~V. Deshmukh, and S.~Nikolaidis, ``Learning from demonstrations using signal temporal logic in stochastic and continuous domains,'' \emph{IEEE Robotics and Automation Letters}, vol.~6, no.~4, pp. 6250--6257, 2021.

\bibitem{hashimoto2022stl2vec}
W.~Hashimoto, K.~Hashimoto, and S.~Takai, ``Stl2vec: Signal temporal logic embeddings for control synthesis with recurrent neural networks,'' \emph{IEEE Robotics and Automation Letters}, vol.~7, no.~2, pp. 5246--5253, 2022.

\bibitem{kmeansalgorithm}
J.~A. Hartigan and M.~A. Wong, ``Algorithm as 136: A k-means clustering algorithm,'' \emph{Journal of the royal statistical society. series c (applied statistics)}, vol.~28, no.~1, pp. 100--108, 1979.

\bibitem{xmeansalgorithm}
D.~Pelleg, A.~Moore \emph{et~al.}, ``X-means: Extending k-means with e cient estimation of the number of clusters,'' in \emph{ICML’00}.\hskip 1em plus 0.5em minus 0.4em\relax Citeseer, 2000, pp. 727--734.

\bibitem{NIPS2017deepsets}
M.~Zaheer, S.~Kottur, S.~Ravanbakhsh, B.~Poczos, R.~R. Salakhutdinov, and A.~J. Smola, ``Deep sets,'' in \emph{Advances in Neural Information Processing Systems}, I.~Guyon, U.~V. Luxburg, S.~Bengio, H.~Wallach, R.~Fergus, S.~Vishwanathan, and R.~Garnett, Eds., vol.~30, 2017.

\bibitem{LSTM}
A.~Graves and A.~Graves, ``Long short-term memory,'' \emph{Supervised sequence labelling with recurrent neural networks}, pp. 37--45, 2012.

\bibitem{paszke2017automatic}
A.~Paszke, S.~Gross, S.~Chintala, G.~Chanan, E.~Yang, Z.~DeVito, Z.~Lin, A.~Desmaison, L.~Antiga, and A.~Lerer, ``Automatic differentiation in pytorch,'' in \emph{NeurIPS Workshop on Autodiff}, 2017.

\bibitem{kingma2014adam}
D.~P. Kingma and J.~Ba, ``Adam: A method for stochastic optimization,'' \emph{arXiv preprint arXiv:1412.6980}, 2014.

\bibitem{pyclustering}
A.~V. Novikov, ``Pyclustering: Data mining library,'' \emph{Journal of Open Source Software}, vol.~4, no.~36, p. 1230, 2019.

\bibitem{andersson2019casadi}
J.~A. Andersson, J.~Gillis, G.~Horn, J.~B. Rawlings, and M.~Diehl, ``Casadi: a software framework for nonlinear optimization and optimal control,'' \emph{Mathematical Programming Computation}, vol.~11, pp. 1--36, 2019.

\bibitem{Calinski–Harabasz}
T.~Caliński and J.~H. and, ``A dendrite method for cluster analysis,'' \emph{Communications in Statistics}, vol.~3, no.~1, pp. 1--27, 1974.

\end{thebibliography}


\begin{thebibliography}{1}
\bibliographystyle{IEEEtran}

\bibitem{ref1}
{\it{Mathematics Into Type}}. American Mathematical Society. [Online]. Available: https://www.ams.org/arc/styleguide/mit-2.pdf

\bibitem{ref2}
T. W. Chaundy, P. R. Barrett and C. Batey, {\it{The Printing of Mathematics}}. London, U.K., Oxford Univ. Press, 1954.

\bibitem{ref3}
F. Mittelbach and M. Goossens, {\it{The \LaTeX Companion}}, 2nd ed. Boston, MA, USA: Pearson, 2004.

\bibitem{ref4}
G. Gr\"atzer, {\it{More Math Into LaTeX}}, New York, NY, USA: Springer, 2007.

\bibitem{ref5}M. Letourneau and J. W. Sharp, {\it{AMS-StyleGuide-online.pdf,}} American Mathematical Society, Providence, RI, USA, [Online]. Available: http://www.ams.org/arc/styleguide/index.html

\bibitem{ref6}
H. Sira-Ramirez, ``On the sliding mode control of nonlinear systems,'' \textit{Syst. Control Lett.}, vol. 19, pp. 303--312, 1992.

\bibitem{ref7}
A. Levant, ``Exact differentiation of signals with unbounded higher derivatives,''  in \textit{Proc. 45th IEEE Conf. Decis.
Control}, San Diego, CA, USA, 2006, pp. 5585--5590. DOI: 10.1109/CDC.2006.377165.

\bibitem{ref8}
M. Fliess, C. Join, and H. Sira-Ramirez, ``Non-linear estimation is easy,'' \textit{Int. J. Model., Ident. Control}, vol. 4, no. 1, pp. 12--27, 2008.

\bibitem{ref9}
R. Ortega, A. Astolfi, G. Bastin, and H. Rodriguez, ``Stabilization of food-chain systems using a port-controlled Hamiltonian description,'' in \textit{Proc. Amer. Control Conf.}, Chicago, IL, USA,
2000, pp. 2245--2249.

\end{thebibliography}

\clearpage
\clearpage

\end{document}